\documentclass[preprint,12pt,authoryear]{elsarticle}

\usepackage{amssymb}
\usepackage{amsmath}

\usepackage{cite}

\usepackage{algorithm}
\usepackage{algorithmicx}
\usepackage{algpseudocode}
\usepackage[ruled, vlined, algo2e]{algorithm2e}
\usepackage{hyperref}
\hypersetup{
hypertex=true,
colorlinks=true,
linkcolor=blue,
anchorcolor=blue,
citecolor=blue
}

\usepackage{multirow}
\usepackage{multicol}
\usepackage{makecell}
\usepackage{booktabs}
\usepackage{adjustbox}
\usepackage{blindtext}
\usepackage{xurl}

\usepackage{lineno}

\journal{Computers and Geotechnics}

\begin{document}
\begin{frontmatter}

\title{Physics-Informed Cross-Learning for Seismic Acoustic Impedance Inversion and Wavelet Extraction} 
\author[label1,label2]{Junheng Peng}
\author[label1,label2]{Xiaowen Wang}
\author[label1,label2]{Yingtian Liu} 
\author[label1,label2]{Yong Li}
\author[label2]{Mingwei Wang}

\affiliation[label1]{organization={The Key Laboratory of Earth Exploration \& Information Techniques of Ministry Education},
            addressline={Chengdu University of Technology}, 
            city={Chengdu},
            postcode={610059}, 
            state={Sichuan},
            country={China}}
\affiliation[label2]{organization={School of Geophysics},
            addressline={Chengdu University of Technology}, 
            city={Chengdu},
            postcode={610059}, 
            state={Sichuan},
            country={China}}

\begin{abstract}

Seismic acoustic impedance inversion is one of the most challenging tasks in geophysical exploration. Many studies have proposed the use of deep learning for processing; however, most of them are limited by factors such as seismic wavelets and low-frequency initial models. Furthermore, self-supervised frameworks constructed entirely using deep learning models struggle to form direct and effective physical constraints to unlabeled outputs during the multi-model concatenation, which leads to instability in inversion. In this work, we introduced innovations in both the deep learning framework and training strategy. First, we designed a deep learning framework to perform acoustic impedance inversion and seismic wavelet extraction simultaneously. Building on this foundation, considering the scarcity of well data, we proposed a physics-informed cross-learning strategy to impose effective constraints on the framework. We conducted comparative experiments and ablation experiments on both synthetic datasets and field datasets. The results demonstrate that the proposed method achieves a significant improvement compared with semi-supervised learning methods and can extract seismic wavelets with relatively high accuracy. Finally, to ensure the reproducibility of this work, we have made the code open-source. 

\end{abstract}

\begin{highlights}
\item We propose a novel deep learning framework for simultaneously acoustic impedance inversion and wavelet extraction.
\item The proposed training strategy incorporates supervised learning, physical constraints, and domain-adaptive learning, without requiring additional seismic wavelets or initial low-frequency models.
\item The proposed method achieves superior performance and robustness compared to previous semi-supervised methods.
\end{highlights}


\end{frontmatter}


\section{Introduction}

The inversion of seismic acoustic impedance is a very important but challenging task in geophysical exploration. What is important is that it can effectively reflect changes in underground lithology and velocity, thus offering an effective reference for research on underground minerals and structures \citep{1, 2}. However, imaging underground acoustic impedance also faces challenges: the cost of a well is expensive (usually a single well costs millions of dollars), and the number of well data limits the accuracy of imaging \citep{3}. The first researchers implemented inversion problems through some heuristic algorithms \citep{4, 5, 6, 7, 8, 9, 10}. These heuristic algorithms include: iterative least squares method \citep{4, 11}, simulated annealing algorithm \citep{5, 12}, ant colony algorithm \citep{6}, genetic algorithm \citep{7}, particle swarm optimization algorithm \citep{8}, and so on. These methods are theoretically well-established \citep{9, 10}; nevertheless, they can hardly achieve the expected results when tackling practical problems. This is because such methods involve an enormous amount of computation and are also plagued by issues such as parameter optimization and initial model establishment.

\begin{figure}
\centering
\noindent\includegraphics[width=\textwidth]{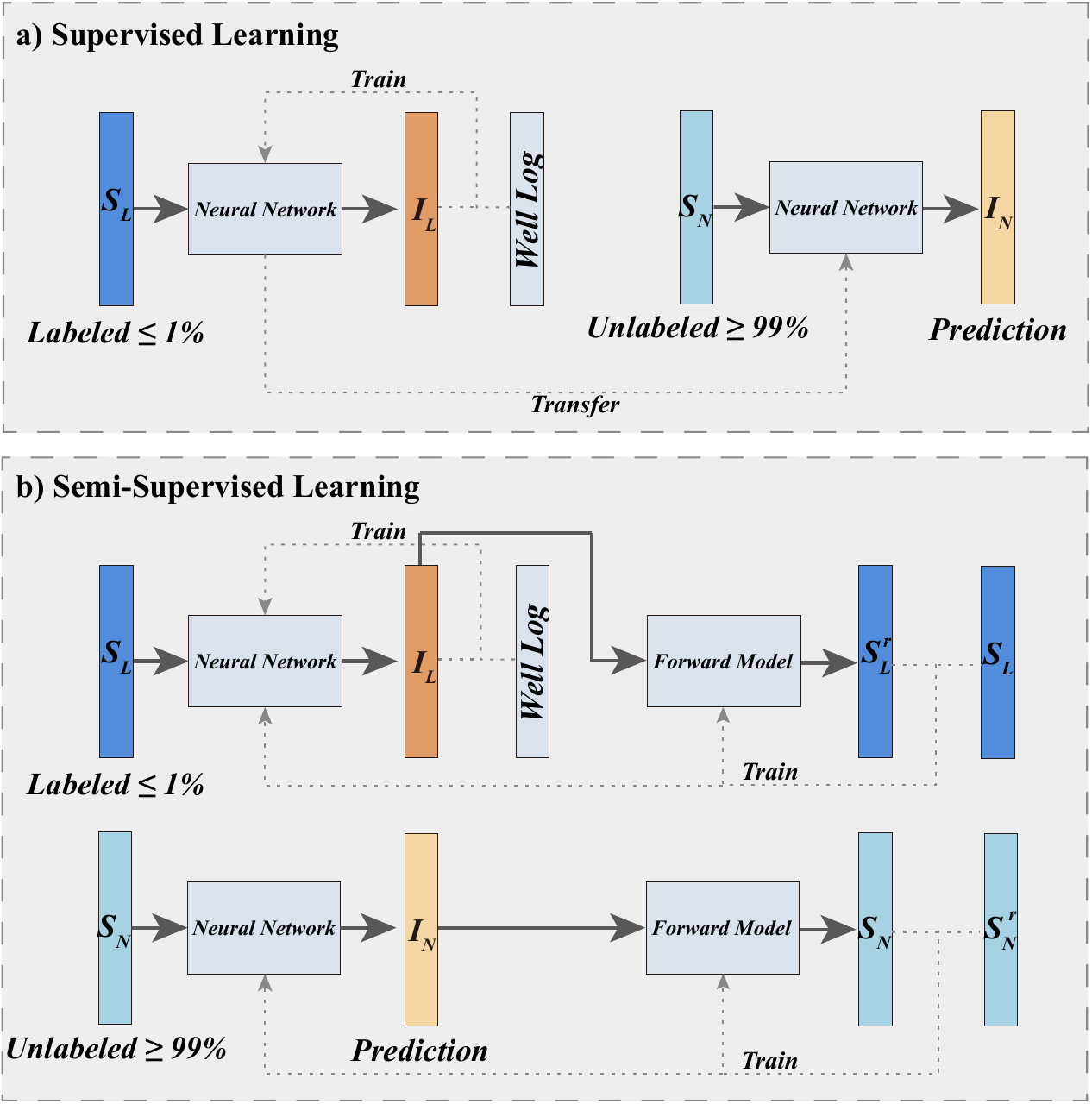}
\caption{Two previous deep learning-based acoustic impedance inversion frameworks.}
\label{framework_compare}
\end{figure}

With the continuous development of artificial intelligence technology, neural networks have influenced various aspects of seismic exploration \citep{23, 24, 51}. The emergence of deep learning has provided a brand-new solution for acoustic impedance inversion \citep{50}. From the perspective of deep learning, the inversion of acoustic impedance is a few-shot problem \citep{13}, due to the very limited number of wells. There are two approaches to impedance inversion using deep learning, as illustrated in Figure~\ref{framework_compare}. On one hand, neural networks struggle to learn well through supervised learning with limited well data, which is why the performance of some early supervised learning methods, such as the convolutional neural network \citep{14} and the temporal convolutional network \citep{15}, could hardly be guaranteed. In recent years, supervised learning has gradually developed along with the evolution of neural network architectures. For example, \citet{16} proposed residual network-based structures, and \citet{17} introduced the self-attention mechanism. On the other hand, to obtain more stable and accurate results, many researchers have proposed semi-supervised learning frameworks \citep{18, 19, 57}. Semi-supervised learning incorporates the physical forward process into the constraints of the inversion network \citep{20, 21}, and can usually achieve more accurate results than supervised learning. This semi-supervised learning method can be further subdivided into two types: one that uses physics-driven forward model, and the other that adopts data-driven neural networks as the forward model. 
 
The method using a physics-driven forward model requires additional approaches to extract seismic wavelets to synthesize seismic data \citep{27}; therefore, the accuracy of such methods depends on the precision of the wavelets. In field exploration, the waveform of seismic wavelets is complex, and the extraction of wavelets is affected by a variety of factors, both of which significantly increase the uncertainty of inversion results. Furthermore, many researchers have proposed that the method using a data-driven forward model avoids this problem effectively \citep{19, 29}. For example, \citet{30} proposed an dual-branch inversion model with attention mechanism to modeling multi-scale sequential features and inversion; \citet{31} proposed a semi-supervised deep learning method with iterative gradient correction for better robustness; \citet{32} developed a fusion neural network that can invert porosity and acoustic impedance simultaneously, which can leverage the patterns learned from similar tasks to enhance the accuracy of impedance inversion; building on the time domain, \citet{33} introduced the frequency domain and Hilbert transform, which enhance the identification of certain weak reflection. It is worth noting that data-driven methods still have certain limitations: such methods usually involve the concatenation of at least two networks, and the simultaneous optimization of these two networks makes it difficult to guarantee the results. Meanwhile, some methods also utilize low-frequency models to enhance inversion accuracy \citep{34}; however, accurate low-frequency models are equally difficult to obtain. 

In this work, we propose a new framework that differs from the aforementioned methods. The method we propose does not require seismic wavelets or forward networks, nor does it rely on low-frequency models or initial models, and thus can effectively alleviate the aforementioned issues. The core idea of the proposed method is to use an encoder to encode seismic data, while employing two parallel networks to simultaneously extract seismic wavelets and invert acoustic impedance. The seismic wavelets and acoustic impedance derived from the proposed method jointly form physics-informed constraints through a physical model, effectively addressing the issues of wavelet acquisition and forward model dependence. Furthermore, since acoustic impedance inversion faces a few-shot learning problem, we propose a cross-learning strategy to enhance the generalization ability when processing unlabeled data and ensure the stability of results. The proposed method does not involve low-frequency model, and its training process exhibits better physical properties compared to deep learning methods with data-driven forward networks. We conducted experiments on three widely used open-source synthetic datasets and one field dataset, and the results show that the proposed method exhibits excellent performance in both computational efficiency and accuracy. Finally, we provide open-source data and code to support the application of deep learning methods in seismic acoustic impedance inversion. 

\section{Method}
\subsection{Physics Modeling}

Acoustic impedance is an important characterization parameter for describing subsurface media, which is widely used in oil, gas, and mineral exploration. Acoustic impedance is defined as:
\begin{equation}
I = V_p \times \rho,
\end{equation}
where $I$ denotes acoustic impedance, $V_p$ denotes the P-wave velocity, and $\rho$ denotes the density of subsurface media. In field exploration, the primary method for measuring acoustic impedance is well logging. However, the cost of a single well is relatively high, which means that obtaining the acoustic impedance of an entire subsurface profile can only be achieved through inversion. The acquisition of seismic data is generally relatively low-cost. After AVO stacking, these records approximate vertical incidence and can reflect the acoustic impedance distribution of subsurface media. Seismic records can be regarded as the convolution of seismic wavelets and reflection coefficient \citep{35}, as shown below: 
\begin{equation}
S(t) = R(t) \ast W(t) + n(t),
\end{equation}
where $\ast$ denotes the convolution operation, $R$ denotes the reflection coefficient, and $W$ denotes the seismic wavelet. $n$ denotes random noise and can be removed using seismic noise attenuation methods \citep{36}. Meanwhile, the reflection coefficient is derived from acoustic impedance:
\begin{equation}
R(t) = \frac{I_{t+\mathrm{d}t} - I_t}{I_{t+\mathrm{d}t} + I_t}, 
\end{equation}
which can defined as the $\Delta$ operator. In summary, we can obtain the equation for synthesizing seismic records from acoustic impedance as:
\begin{equation}
S(t) = [\Delta \cdot I(t)]\ast W(t) + n(t).
\label{seismic_synthetic}
\end{equation}
It is worth noting that when the seismic wavelet and acoustic impedance are known, the forward synthesis of seismic records is relatively straightforward; however, when only seismic data and a small amount of well logging data are available, it is extremely difficult to obtain acoustic impedance through inversion due to the band-limited characteristic of wavelet. 

\subsection{Structure of Models}

Most works on semi-supervised acoustic impedance inversion first extract wavelets using external algorithms, and their results are largely affected by the accuracy of the wavelets. However, it can be seen from Equation~\ref{seismic_synthetic} that seismic wavelets and acoustic impedance are of equal importance, and the process of extracting seismic wavelets is highly correlated with acoustic impedance inversion. Direct and simultaneous optimization of the two constitutes a non-convex optimization problem. Traditional solutions involve heuristic algorithms, which are typically computationally expensive and rely on a well-designed initial model. 

\begin{figure}
\centering
\noindent\includegraphics[width=\textwidth]{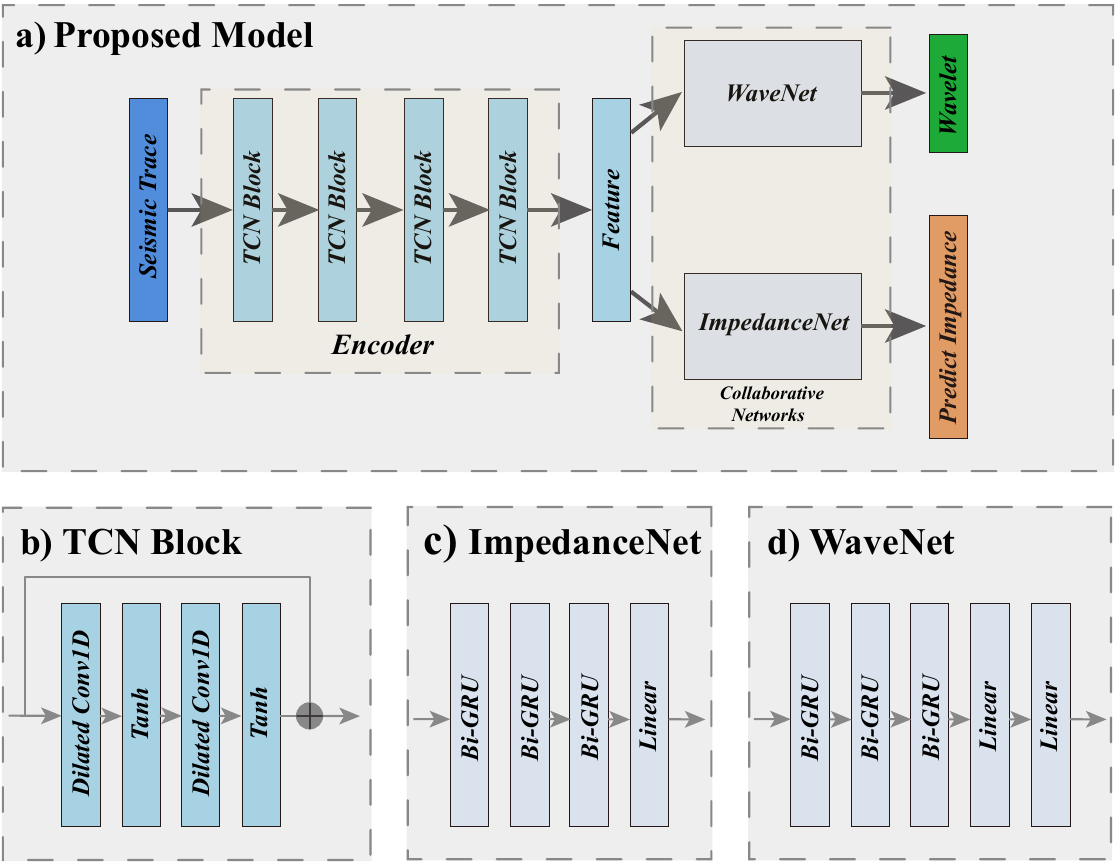}
\caption{a) the structure of proposed network; b) the structure of TCN block; c) the structure of ImpedanceNet; d) the structure of WaveNet.}
\label{model_structure}
\end{figure}

Therefore, in this work, we design a neural network for the simultaneous extraction of seismic wavelets and acoustic impedance inversion, as shown in Figure~\ref{model_structure}. The proposed model is mainly composed of an encoder and two collaborative networks: the encoder extracts features from the input seismic data, while the two collaborative networks are responsible for wavelet extraction and impedance inversion, respectively. The encoder is constructed using the widely adopted temporal convolutional network (TCN) blocks \citep{38}; each TCN block (Figure~\ref{model_structure}b) consists of two dilated convolutional layers and activation function layer. Meanwhile, it is worth noting that the forward process of seismic data is a non-causal process, which has been validated by many studies \citep{37}. Therefore, we adopt a non-causal TCN, which abandons the causal property and is constructed solely using dilated convolution. Benefiting from dilated convolution, the TCN block is capable of handling long sequences effectively. In this work, the encoder is composed of four TCN blocks, and the channels of their features is 16, 16, 16, and 32, respectively. For each TCN block, the dilation factor of the convolutional layers is set to 2, and the Tanh function is used as the activation function. 

In addition, the two collaborative networks for downstream tasks adopt different structures with the encoder, as shown in Figure~\ref{model_structure}c and d. Compared with non-causal TCN, another widely used network is the gated recurrent unit (GRU) or bidirectional GRU (Bi-GRU) \citep{39, 40}. GRU is a variant of recurrent neural networks (RNNs). It improves RNNs through gating units, and a single layer of GRU is sufficient to complete a full sequence modeling process. In this work, we used Bi-GRU to construct ImpedanceNet, which consists of three Bi-GRU layers, and the number of feature channels in each Bi-GRU layer is 32. Finally, we add a linear layer for channel reduction. Furthermore, as we described, the two tasks of wavelet extraction and impedance inversion are highly relevant and similar. Therefore, for WaveNet used for wavelet extraction, we adopt the similar structure as that of ImpedanceNet. The difference is that we added an additional linear layer to the WaveNet for scaling the length of the wavelet. 

\subsection{Physics-Informed Cross-Learning}

\begin{figure}
\centering
\noindent\includegraphics[width=\textwidth]{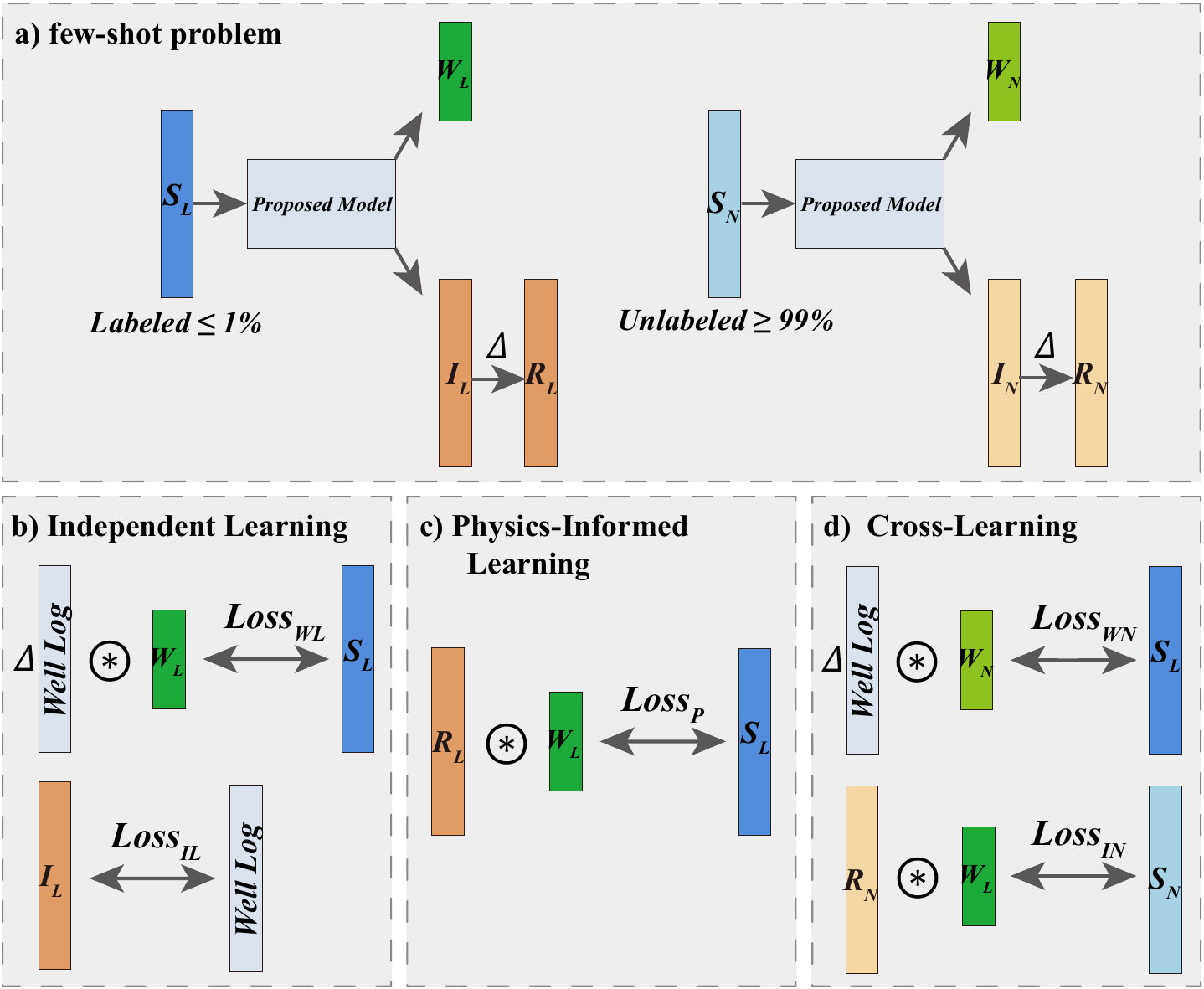}
\caption{a) the few shot problem of acoustic impedance inversion; b) the independent learning of the ImpedanceNet and WaveNet; c) the physics-informed learning on $S_L$; d) the cross-learning between $S_L$ and $S_N$.}
\label{cross_learning}
\end{figure}

Since we adopt a network different from that of semi-supervised methods and there is no involvement of a forward network, the semi-supervised learning strategy is not applicable to the proposed method. On this basis, we propose a new learning strategy that can learn both labeled and unlabeled datasets simultaneously, as shown in Figure~\ref{cross_learning}, which named physics-informed cross-learning. Typically, the fully acquired $N$ seismic traces $S=\{s_x\}_{x=1}^N$ consist of labeled traces $S_L=\{s_l\}_{l=1}^{N_L}$ and unlabeled traces $S_N=\{s_n\}_{n=1}^{N_N}$, where $N_L+N_N = N$ and $N_L\ll N_N$. The impedance dataset $I_L = \{i_l\}_{l=1}^{N_L}$ corresponding to the labeled traces $S_L$ is mainly derived from well-logging, and their relationship is expressed as $S_L = [\Delta \cdot I_L]\ast W$, where we assume that the noise has been attenuated. The goal of acoustic impedance inversion is to obtain $I_N=\{i_n\}_{n=1}^{N_N}$ in $S_N = [\Delta \cdot I_N]\ast W$ when $S$ and $I_L$ are known. The idea of supervised learning is to train a deep learning model using $S_L$ and $I_L$, and directly transfer it to $S_N$ for impedance inversion. However, such methods usually struggle to achieve stable results, as a small training set tends to cause the network to suffer from overfitting. Physics-informed semi-supervised learning takes into account the forward process of seismic data, but it faces the challenge of wavelet extraction; in contrast, data-driven semi-supervised learning struggles to ensure that the learning results of the forward network and the inversion network align with the target. To address these issues, we improved the structure of the deep learning network and proposed a physics-informed cross-learning strategy. 

The proposed strategy consists of three components: independent learning, physics-informed learning, and cross-learning, each playing a crucial role. First, regarding the independent learning component, the design objective of this component is to train the ImpedanceNet and WaveNet using existing labeled data, which is also of great importance in the supervised and semi-supervised learning strategies. The loss of this component can be expressed as:
\begin{equation}
\begin{split}
Loss_{I} &= Loss_{WL} + Loss_{IL} \\ &= \| [\Delta \cdot i_l] \ast WaveNet(E(s_l)) -s_l \|^2_2 + \| ImpNet(E(s_l))-i_l \|^2_2,
\end{split}
\label{loss_I}
\end{equation}
where only the labeled datasets $S_L$ and $I_L$ are used in this loss, ensuring that the network learning results conform to field conditions. In addition, considering the forward process in Equation~\ref{seismic_synthetic}, we proposed physics-informed learning component, as shown in Figure~\ref{cross_learning}c. Unlike the independent learning component, which optimizes the ImpedanceNet and WaveNet separately, the physics-informed learning component establishes mutual constraints between the ImpedanceNet and WaveNet by leveraging the physical forward process. The loss function for physics-informed learning can be expressed as:
\begin{equation}
Loss_{P} = \| [\Delta \cdot ImpNet(E(s_l))] \ast WaveNet(E(s_l)) -s_l \|^2_2.
\label{loss_P}
\end{equation}
Similarly, the aforementioned learning operates on the labeled dataset. In order to ensure the robustness of the networks on the unlabeled dataset, we proposed cross-learning component. Unlike the other two components, the purpose of cross-learning is to reduce the error caused by the discrepancy between the labeled dataset $S_L$ and the unlabeled dataset $S_N$. Such a problem is generally referred to as open-set domain adaptation \citep{41}, and the common solution is to perform feature alignment in the feature space \citep{42, 43}. The proposed cross-learning is essentially a physics-informed semi-supervised feature alignment technology. It aligns the features of labeled and unlabeled data using equations as constraints, which can be expressed as:
\begin{equation}
\begin{split}
Loss_{C} &= Loss_{WN} + Loss_{IN} \\ &= \| [\Delta \cdot i_l] \ast WaveNet(E(s_n)) -s_l \|^2_2 \\ &+ \| [\Delta \cdot ImpNet(E(s_n))] \ast WaveNet(E(s_l)) - s_n \|^2_2.
\end{split}
\label{loss_C}
\end{equation}
During training, labeled and unlabeled data are randomly sampled with the same batch size as 6. Looking back at the deep learning model structure in previous section, it is closely associated with the three learning strategies we put forward. When only $Loss_I$ is used, the proposed method is equivalent to ordinary supervised learning; the integration of WaveNet enables the calculation of $Loss_P$ and $Loss_C$, where $Loss_P$ implements physics-informed constraint and $Loss_C$ ensures domain adaptation for unlabeled data. Among these, each of the three loss functions includes the optimization for both ImpedanceNet and WaveNet, respectively. 

\begin{figure}
\centering
\noindent\includegraphics[width=\textwidth]{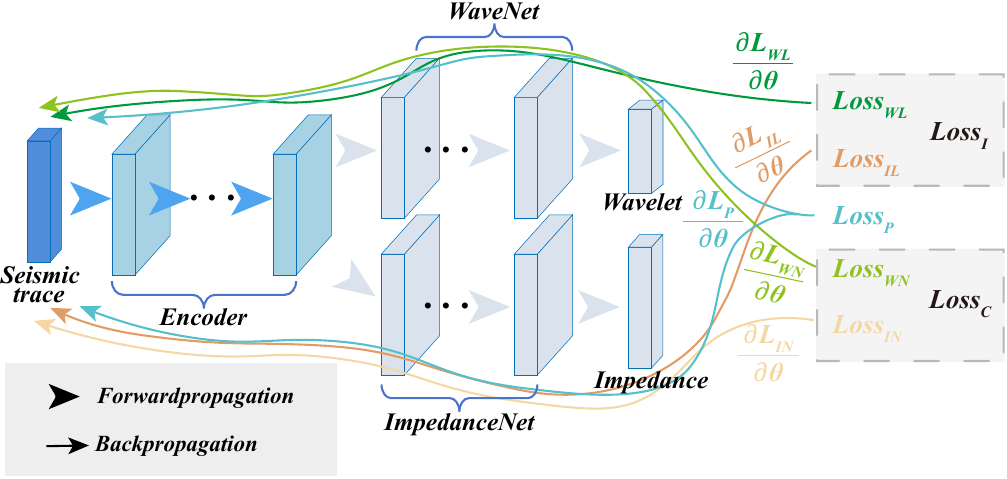}
\caption{The forward propagation and backpropagation of proposed method.}
\label{backpropagation}
\end{figure}

In addition, to ensure the stability of network training, we have optimized the backpropagation paths of some losses, as illustrated in Figure~\ref{backpropagation}. The backpropagation path simultaneously takes into account the independent learning, physics-informed learning, and cross-learning of both the WaveNet and ImpedanceNet.

\subsection{Inversion Workflow}

In summary, the overall processing workflow of the proposed method is shown in Algorithm~\ref{workflow}. First, random data are sampled separately from the labeled and unlabeled datasets with the same batch size, and the processed results are obtained through forward propagation. On this basis, we calculated the losses for independent learning, physics-informed learning, and cross-learning respectively. Finally, gradient descent is performed on the losses to optimize the parameters of the neural networks. During this process, the AdamW optimizer was employed for parameter learning, with an initial learning rate of 0.004. To ensure the convergence of network parameters, the total number of training epochs was set to 1000. Furthermore, a learning rate decay strategy was adopted, where the learning rate was reduced by 50$\%$ every 250 training epochs.
\begin{algorithm}
        \setcounter{AlgoLine}{0}
        \scriptsize
        \LinesNumbered
        \caption{The workflow of proposed method.}\label{workflow}
        \KwIn{Seismic traces $S = S_L \cup S_N$; Few seismic traces $S_L$ and matching well-logging data $I_L$; Unlabeled seismic traces $S_N$; Initialized neural networks $E$, $WaveNet$, and $ImpNet$}
        \KwOut{Acoustic Impedance $I = I_L \cup I_N $}
        \tcp{training}
        \;
        \For{$i=1$ to $Max\ Epoch$}{
        \For{$j=1$ to $Max\ Batch$}{
        
        Randomly extract matching $s_l$ and $i_l$ from $S_L$ and $I_L$ 

        Randomly extract $s_n$ from $S_N$
        
        Get the output of $s_l$: $\hat{w_l} \leftarrow WaveNet(E(s_l|\theta_E)| \theta_W)$, $\hat{i_l} \leftarrow ImpNet(E(s_l|\theta_E)| \theta_I)$ 

        Get the output of $s_n$: $\hat{w_n} \leftarrow WaveNet(E(s_n|\theta_E)| \theta_W)$, $\hat{i_n} \leftarrow ImpNet(E(s_n|\theta_E)| \theta_I)$
        
        \tcp{Calculate the losses}
        
        Independent learning: $Loss_{I} \leftarrow \| [\Delta \cdot i_l] \ast \hat{w_l} -s_l \|^2_2 + \| \hat{i_l}-i_l \|^2_2$ 

        Physics-informed learning: $Loss_{P} \leftarrow \| [\Delta \cdot \hat{i_l}] \ast \hat{w_l} -s_l \|^2_2$

        Cross-learning: $Loss_{C} \leftarrow \| [\Delta \cdot i_l] \ast \hat{w_n} -s_l \|^2_2 + \| [\Delta \cdot \hat{i_n}] \ast detach(\hat{w_l}) - s_n \|^2_2$
        
        \tcp{backpropagation}
        
        $\theta_E, \theta_W, \theta_I \leftarrow \mathop{\arg\min}\limits_{\theta_E, \theta_W, \theta_I}(Loss_I+Loss_P+Loss_C)$
        }}
        
        \tcp{Inversion}
        \For{$s_i$ in $S$}{
        $i_i \leftarrow ImpNet(E(s_i|\theta_E)| \theta_I)$
       }
        \textbf{Return:} inverted acoustic impedance $I = \{i_i\}_{i=1}^N$
\end{algorithm}

After the training is completed, the trained neural network is applied to all seismic traces, and the inversion results of acoustic impedance can thus be obtained. Throughout the entire process, the hardware environment we used consisted of an Intel Core i5-14400F CPU with 16GB memory and an NVIDIA RTX 4060 GPU with 8GB memory; The software environment included CUDA 11.8, PyTorch 2.4.0, and Python 3.8.20. In the next section, we will conduct evaluation experiments using synthetic data and field data to verify the effectiveness and efficiency of the proposed method. Furthermore, to ensure the reproducibility of the results, we have made the data and code open-source.

\section{Examples}
\subsection{Synthetic Datasets}
\subsubsection{Datasets}
In this section, we evaluated the performance of our proposed method using three extensively adopted benchmark datasets: Overthrust \citep{45}, Marmousi 2 \citep{44}, and SEAM elastic earth model \citep{46}. Over the past decade, these datasets have established themselves as standard evaluation benchmarks in acoustic impedance inversion research, making them ideal for comparing the effectiveness of our method against existing approaches. The key specifications for each dataset are compiled in Table~\ref{datasets} for clarity. A critical design choice in our experiments is the use of labeled data: we employ less than 1$\%$ of the total seismic traces (paired with the corresponding well-logging data) as labeled samples. This proportion is substantially lower than the labeled data ratios commonly reported in prior studies, which typically rely on more annotated samples to train inversion models. Notably, the Marmousi 2 dataset poses unique challenges due to its higher structural complexity and greater inversion difficulty compared to the other two datasets. To address this complexity while maintaining the parsimony of labeled data, we use a marginally larger number of labeled traces for Marmousi 2. Even so, this adjusted number remains smaller than the labeled sample sizes used for Marmousi 2 in most previous studies. Furthermore, to ensure the validity and independence of experimental results, each dataset is utilized separately for validation, without data overlap or sharing across different datasets. Furthermore, the deep learning hyperparameters used in all experiments were kept consistent to ensure the objectivity of the experiments.
\begin{table}[!htbp]
\centering
\setlength{\tabcolsep}{1.5mm}
\renewcommand\arraystretch{1.2}
\begin{tabular}{ccccc}
\toprule[1.5pt]
\textbf{Dataset} & \textbf{Trace Num}  & \textbf{Well Num} & \textbf{Time Interval} & \textbf{Distance} \\
\midrule
Overthrust & 1067 & 10 $(\textless 1\%)$ & 4ms & 10.7km \\
\midrule
Marmousi 2 & 1701 & 14 $(\textless 1\%)$ & 3ms & 13.6km \\
\midrule
SEAM & 1751 & 10 $(\textless 1\%)$ & 4ms & 35km \\
\bottomrule[1.5pt]
\end{tabular}
\caption{The parameters of Overthrust, Marmousi 2 and SEAM datasets}
\label{datasets}
\end{table}

\begin{figure}
\centering
\noindent\includegraphics[width=\textwidth]{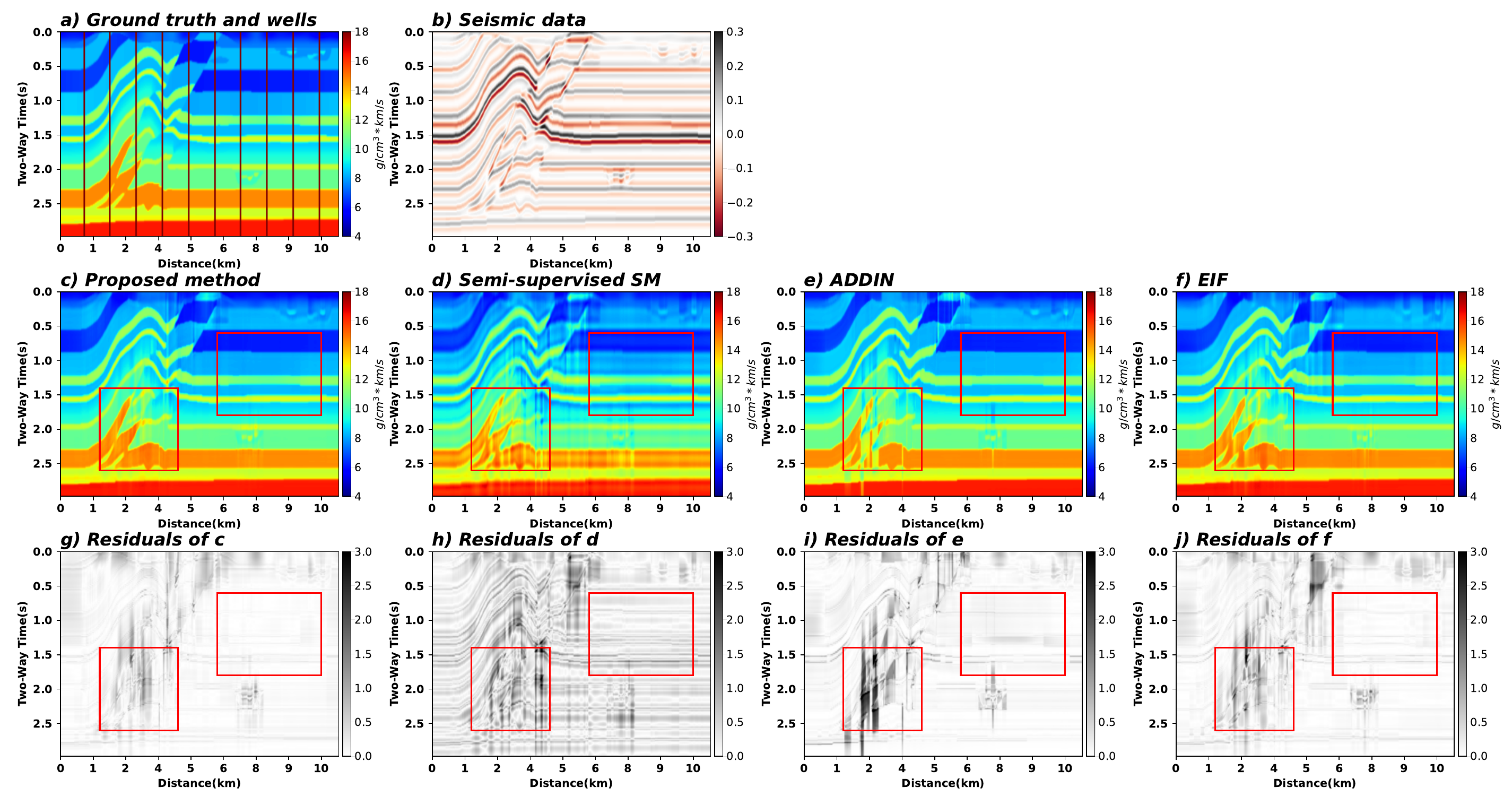}
\caption{The seismic data, true acoustic impedance and results of Overthrust. a) represents Overthrust acoustic impedance and wells; b) represents the seismic data; c) to f) represent the results of proposed method, semi-supervised SM, ADDIN, and EIF respectively; g) to j) represent the absolute residuals between c) to f) and ground truth, respectively.}
\label{overthrust_whole}
\end{figure}

The first dataset is Overthrust, which is developed by the SEG/EAGE 3-D Modeling Committee \citep{45} and stands as a relatively uncomplicated yet widely utilized benchmark. For reference, Figures~\ref{overthrust_whole}a and b present the acoustic impedance and seismic data of the Overthrust, respectively. To compare proposed method with existing techniques, we selected three open-source methods for comparative analysis: a semi-supervised sequence modeling method integrating physics-driven forward process (semi-supervised SM) \citep{18}, an attention-based dual-branch double-inversion network with model-driven forward modeling (ADDIN) \citep{30}, and an encoder-inverter framework without forward process (EIF) \citep{47}. In line with our experimental design, we selected 10 seismic traces at equal spatial intervals to serve as labeled samples, which are marked in Figure~\ref{overthrust_whole}a. In Figure~\ref{overthrust_whole}, it can be clearly observed that the acoustic impedance obtained by the proposed method is the closest to the ground truth. Specifically, as shown in the red box, it accurately reproduces geological features, such as the continuity of impedance interfaces, the magnitude of impedance variations (ranging approximately from 1.0 $g/cm^3 * km/s$ to 2.5 $g/cm^3 * km/s$), and the spatial distribution of high and low impedance zones. In contrast, the semi-supervised SM exhibits slight blurring at the boundaries of impedance interfaces (Figure~\ref{overthrust_whole}d), which adversely affects the judgment of reservoir or formation boundaries. ADDIN shows insufficient performance in handling complex structures ((Figure~\ref{overthrust_whole}e), while the EIF is significantly inferior to the proposed method in terms of lateral continuity (Figure~\ref{overthrust_whole}f). The residual maps (Figures~\ref{overthrust_whole}g to j) reflect the absolute differences between each method’s result and the ground truth (Figure~\ref{overthrust_whole}a). The residual map of the proposed method (Figure~\ref{overthrust_whole}g) exhibits the lowest overall residual amplitude (mostly concentrated near 0 to 0.5) and the fewest high-residual anomalies, indicating minimal deviation from the ground truth. In contrast, the residual maps of semi-supervised SM, ADDIN, and EIF (Figures~\ref{overthrust_whole}h to j) all show more widespread high-residual regions (with local values reaching 2.5 to 3.0) that further confirms that the proposed method achieves more stable inversion result.  

\begin{figure}
\centering
\noindent\includegraphics[width=\textwidth]{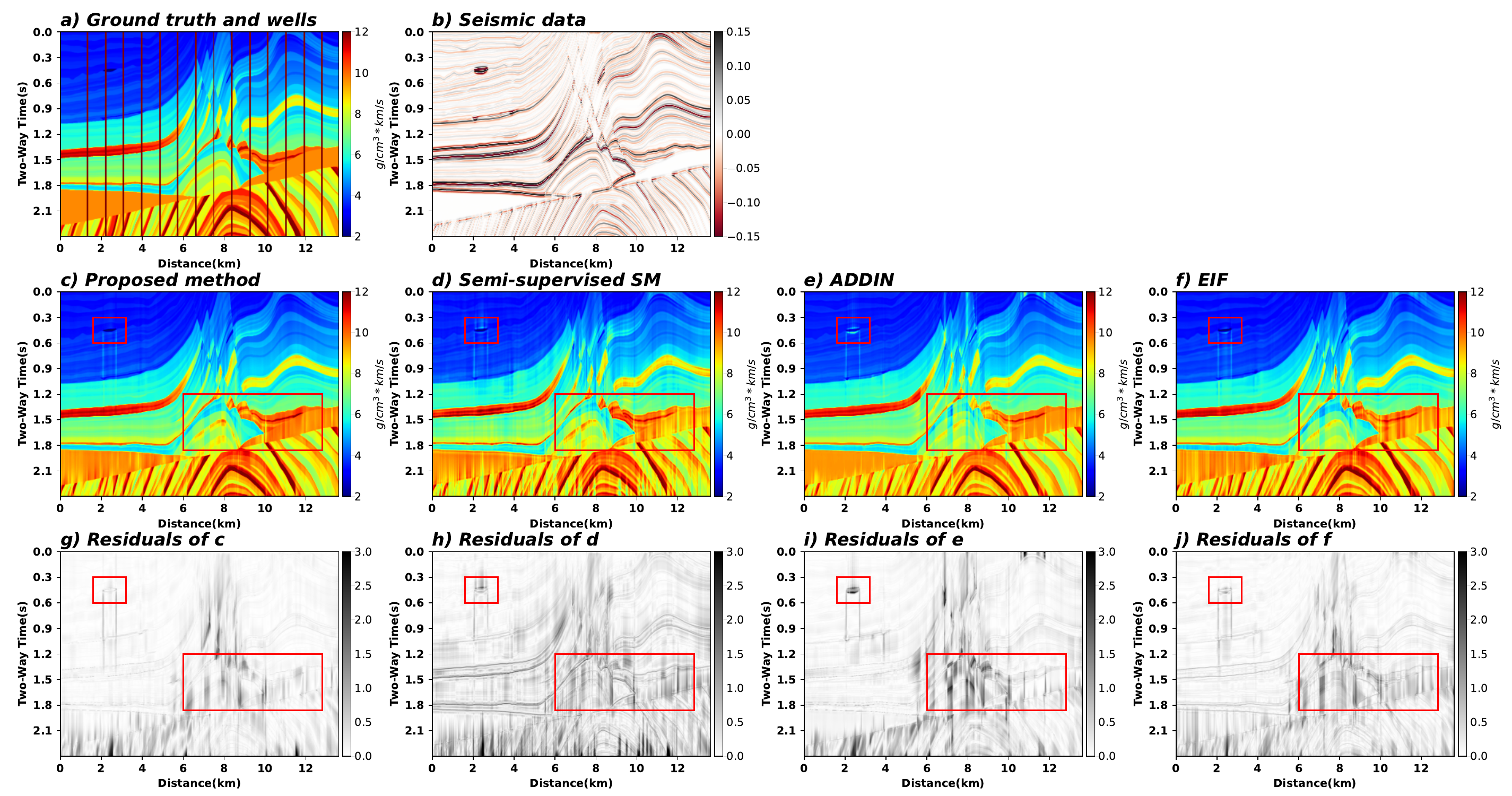}
\caption{The seismic data, true acoustic impedance and results of Marmousi 2. a) represents Overthrust acoustic impedance and wells; b) represents the seismic data; c) to f) represent the results of proposed method, semi-supervised SM, ADDIN, and EIF respectively; g) to j) represent the absolute residuals between c) to f) and ground truth, respectively.}
\label{marmousi_whole}
\end{figure}

The second dataset is the AGL elastic Marmousi, commonly referred to as Marmousi 2 (Figures~\ref{marmousi_whole}a and b), was developed by \citet{44}. Compared with the Overthrust dataset, the Marmousi 2 dataset features more complex geological structures and distinct anisotropy, making it widely used for validating acoustic impedance inversion workflows. Figures~\ref{marmousi_whole}c to f present the inversion results of four methods. Visually, the inversion result of the proposed method (Figure~\ref{marmousi_whole}c) exhibits the highest consistency with the ground truth, which accurately reproduces the continuous distribution of impedance interfaces and the variations in acoustic impedance. Similar to the results from the Overthrust dataset, the semi-supervised SM (Figure~\ref{marmousi_whole}d) exhibits significant errors at certain impedance boundaries; the ADDIN (Figure~\ref{marmousi_whole}e) shows insufficient resolution for complex and fine-scale structures; and the EIF (Figure~\ref{marmousi_whole}f) yields results with poor continuity. These observations are also evident in the residual maps, particularly in the regions marked by the red boxes in Figures~\ref{marmousi_whole}g to j. The residual map of the proposed method (Figure~\ref{marmousi_whole}g) exhibits the lowest overall residual amplitude (mostly concentrated around 0.5), with high-residual regions primarily confined to complex fault zones. In contrast, the residual maps of semi-supervised SM (Figure~\ref{marmousi_whole}h), ADDIN (Figure~\ref{marmousi_whole}i), and EIF (Figure~\ref{marmousi_whole}j) all show more extensive high-residual areas. 

\begin{figure}
\centering
\noindent\includegraphics[width=\textwidth]{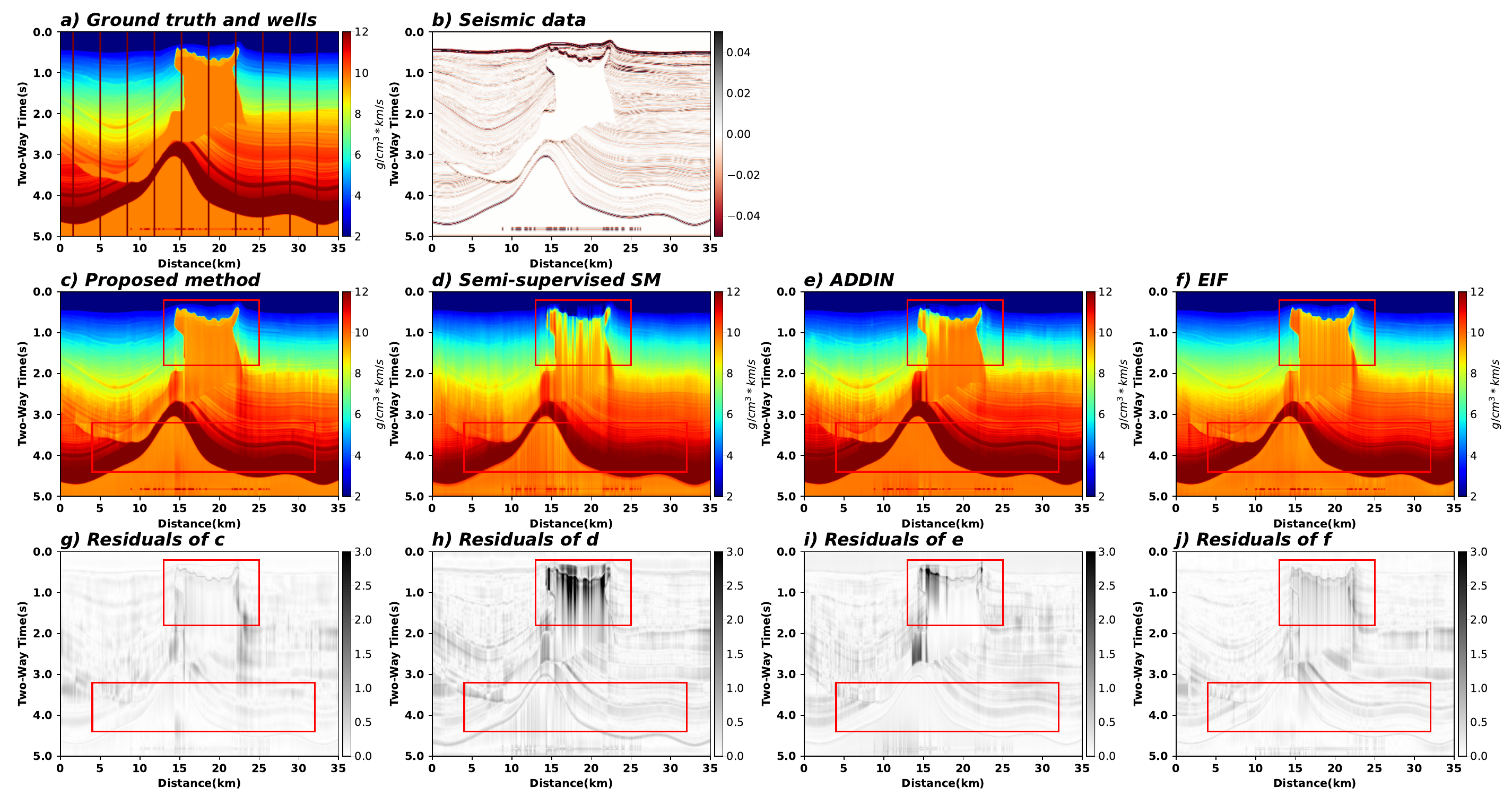}
\caption{The seismic data, true acoustic impedance and results of SEAM. a) represents Overthrust acoustic impedance and wells; b) represents the seismic data; c) to f) represent the results of proposed method, semi-supervised SM, ADDIN, and EIF respectively; g) to j) represent the absolute residuals between c) to f) and ground truth, respectively.}
\label{seam_whole}
\end{figure}

The last dataset is the SEAM elastic earth model \citep{46}. While SEAM is less geologically intricate compared to Marmousi 2, it has a significantly larger spatial extent, covering around 35km, and all its properties are derived from basic rock properties. Besides containing a sandstone body, SEAM also features abundant thin layers, as shown in Figures~\ref{seam_whole}a and b. Figures~\ref{seam_whole}c to f present the inversion results of the four methods. It can be seen that the inversion result of the proposed method (Figure~\ref{seam_whole}c) exhibits the highest consistency with the ground truth and it accurately reproduces the continuous spatial distribution of impedance interfaces across the entire profile, particularly the intermediate sandstone. For comparison, the semi-supervised SM (Figure~\ref{seam_whole}d) shows significant blurring at the boundaries between the sandstone and low-impedance layers, accompanied by obvious loss of thin layers in the mid-to-deep profile. The ADDIN and the EIF (Figure~\ref{seam_whole}f) similarly struggle with these issues, displaying distinct discontinuities at the thin layers and around the sandstone, which indicates their limited ability to capture fine-scale geological features and maintain structural continuity in impedance inversion. Furthermore, the residual maps in Figures~\ref{seam_whole}g and j also clearly show that the errors of the three comparative methods are mainly concentrated in the thin-layer and sandstone regions.

\subsubsection{Quantitative Evaluation}

\begin{figure}
\centering
\noindent\includegraphics[width=\textwidth]{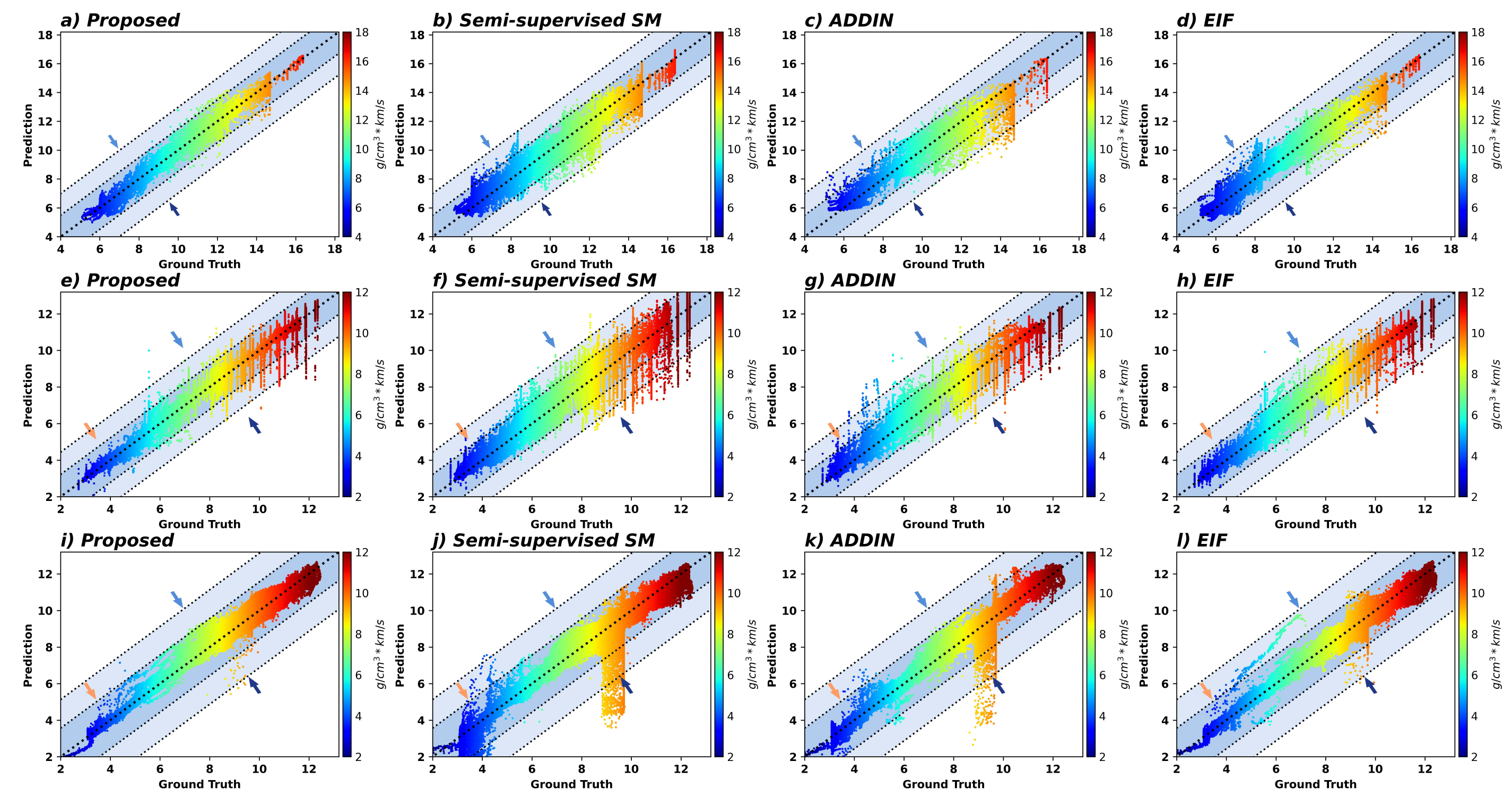}
\caption{The scatterplots of results and true acoustic impedance on each dataset. a) to d) represent the results of each method for Overthrust; e) to h) represent the results of each method for Marmousi 2; i) to l) represent the results of each method for SEAM.}
\label{synthetic_scatter}
\end{figure}

\begin{figure}
\centering
\noindent\includegraphics[width=\textwidth]{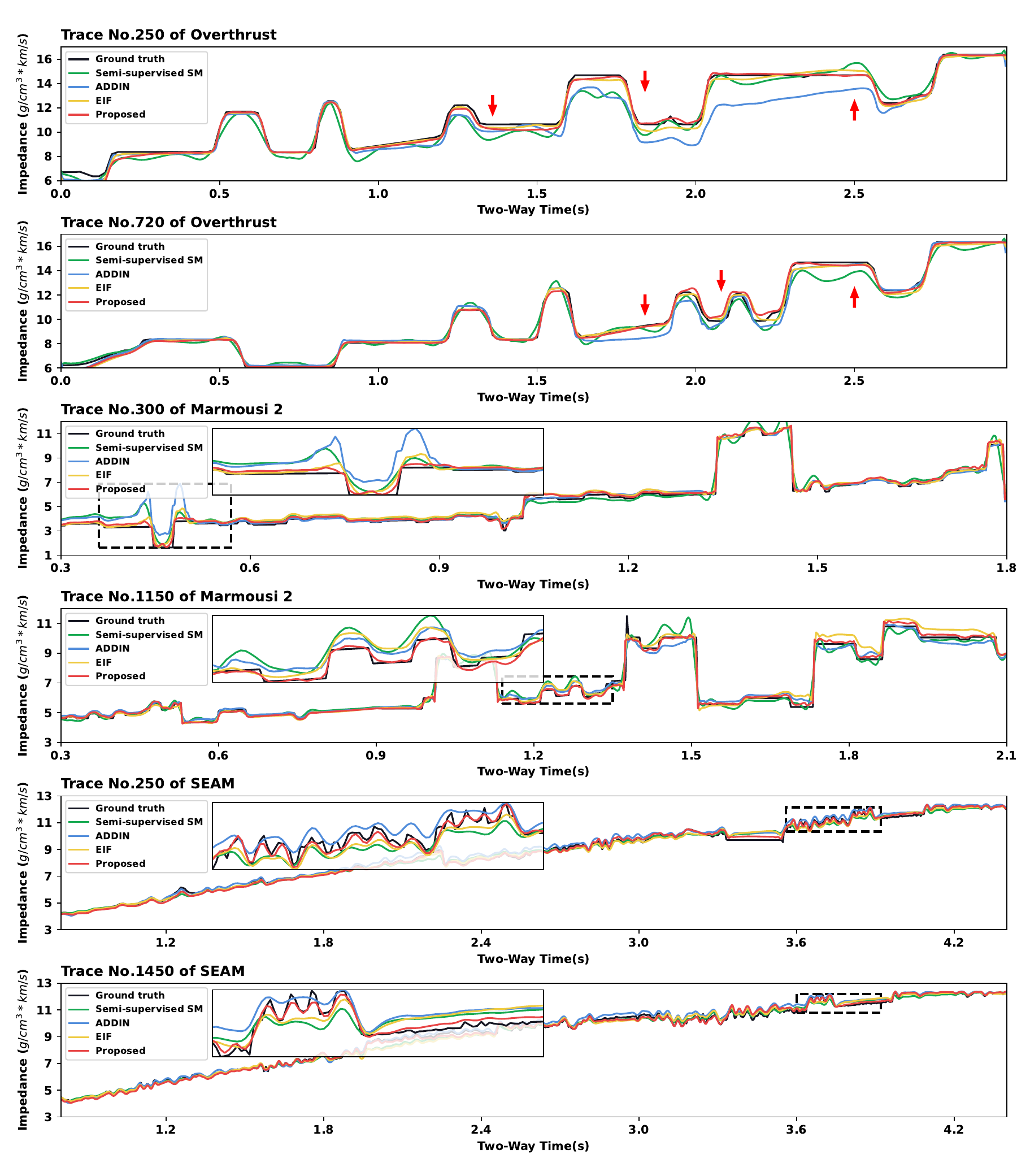}
\caption{The comparison of selected traces on each dataset. }
\label{synthetic_trace}
\end{figure}

To more intuitively observe the errors between the results of various methods and the true acoustic impedance, we plotted scatter maps of each inversion result against the ground truth (Figure~\ref{synthetic_scatter}), and several seismic traces selected for visualization (Figure~\ref{synthetic_trace}). From the perspective of the scatter maps, the scatter points of the proposed method cluster most tightly around the 45 degree diagonal line, which indicates a strong linear relationship between its inversion results and the ground truth (Figures~\ref{synthetic_scatter}a, e, and i). In contrast, the semi-supervised SM exhibits noticeable deviations, particularly when the acoustic impedance values fall within the range of 3 to 8 $g/cm³*km/s$ (Figure~\ref{synthetic_scatter}j, a range that corresponds to the sandstone body and thin-layer regions of SEAM). While the scatter points of the ADDIN and EIF are relatively concentrated overall, there remain a small number of outlier points with significant deviations in specific value ranges. This pattern of error distribution aligns well with the observations from the residual maps in the preceding section, where these two methods also showed localized high-residual regions. Figure \ref{synthetic_trace} further zooms in to present the results of each method along 6 selected seismic traces. It is distinctly observable that the curve of the proposed method exhibits the highest degree of overlap with the ground truth, and it accurately captures the subtle peaks and troughs of impedance in thin layers (for instance, the shallow reservoirs in the Marmousi 2 dataset and the weak reflection zones in the SEAM), which are fine-scale geological features mentioned in previous section. In contrast, the other comparative methods show a phenomenon of peak flattening in thin-layer regions. This flattening effect prevents them from resolving fine-scale impedance variations, thereby failing to reproduce the detailed geological heterogeneities, which is consistent with the earlier observations of thin-layer loss and boundary blurring in their inversion result. 

Furthermore, five metrics are used to perform precise comparative experiments. The first metric is the signal-to-noise ratio (SNR). Widely utilized in signal processing, it is defined as:
\begin{equation}
SNR = 10\times\log_{10}\frac{\sum_{i = 1}^{N}I_{i}^{2}}{\sum_{i = 1}^{N}(I_{i}-I_{i}^{\prime})^{2}}
\label{equation: SNR}
\end{equation}
where $I$ represents the true acoustic impedance, and $I^{\prime}$ represents the predicted acoustic impedance. SNR characterizes the logarithmic ratio between the true impedance and the prediction error. Higher SNR values signify superior performance. The second metric is $R^{2}$, the most prevalently used evaluation metric in AII. It is defined as:
\begin{equation}
R^{2}=1 - \frac{\sum_{i = 1}^{N}(I_{i}-I_{i}^{\prime})^{2}}{\sum_{i = 1}^{N}(\mu_{I}-I_{i})^{2}},
\label{equation: R2}
\end{equation}
where $\mu_{I}$ is the mean of $I$. Structural similarity (SSIM) is a metric extensively applied in image processing to depict the similarity between two images. It has also been adopted in numerous AII research works. Its definition is given by:
\begin{equation}
SSIM=\frac{(2\times\mu_{I}\times\mu_{I^{\prime}}+c_{1})(2\times\sigma_{II^{\prime}}+c_{2})}{(\mu_{I}^{2}+\mu_{I^{\prime}}^{2}+c_{1})(\sigma_{I}^{2}+\sigma_{I^{\prime}}^{2}+c_{2})},
\label{equation: SSIM}
\end{equation}
where $\sigma_{I}^{2}$ represents the variance of $I$, $\sigma_{I^{\prime}}^{2}$ denotes the variance of $I^{\prime}$, and $\sigma_{II^{\prime}}$ stands for the covariance between $I$ and $I^{\prime}$. The parameters $c_{1}$ and $c_{2}$ are small positive constants to prevent division by zero. The Structural Similarity Index Measure (SSIM) demonstrates a higher sensitivity to global deviations and is less affected by local errors. Mean Absolute Error (MAE) and Mean Squared Error (MSE) are two metrics that are widely utilized in regression tasks. Their definitions are as follows:
\begin{equation}
MAE=\frac{1}{N}\sum_{i = 1}^{N}|I_{i}-I_{i}^{\prime}|,
\label{equation: MAE}
\end{equation}
\begin{equation}
MSE=\frac{1}{N}\sum_{i = 1}^{N}(I_{i}-I_{i}^{\prime})^{2}.
\label{equation: MAE}
\end{equation}
\begin{table}[!htbp]
\centering
\tiny
\setlength{\tabcolsep}{3mm}
\renewcommand\arraystretch{1.6}
\begin{tabular}{cccccc}
\toprule[1.5pt]
\textbf{Metric} & \textbf{Dataset}  & \textbf{Proposed} & \textbf{\makecell{Semi-supervised \\ SM}} & \textbf{ADDIN} & \textbf{EIF} \\
\midrule
\multirow{3}{*}{\textbf{SNR $\uparrow $}} & Overthrust & \textbf{33.4297} & 27.0136 & 28.9334 & 30.8307 \\
& Marmousi 2 & \textbf{27.0205} & 22.3415 & 24.3346 & 25.7436 \\
& SEAM & \textbf{32.6203} & 25.3898 & 29.0701 & 31.5077 \\
\midrule
\multirow{3}{*}{\textbf{$R^2 \uparrow$}} & Overthrust & \textbf{0.9941} & 0.9740 & 0.9833 & 0.9892 \\
& Marmousi 2 & \textbf{0.9838} & 0.9524 & 0.9699 & 0.9783 \\
& SEAM & \textbf{0.9956} & 0.9769 & 0.9901 & 0.9944 \\
\midrule
\multirow{3}{*}{\textbf{SSIM $\uparrow $}} & Overthrust & \textbf{0.9728} & 0.8809 & 0.9111 & 0.9542\\
& Marmousi 2 & \textbf{0.9409} & 0.8355 & 0.8948 & 0.9134 \\
& SEAM & \textbf{0.9510} & 0.8711 & 0.9124 & 0.9277 \\
\midrule
\multirow{3}{*}{\textbf{MAE $\downarrow$}} & Overthrust & \textbf{0.0443} & 0.1126 & 0.0636 & 0.0579\\
& Marmousi 2 & \textbf{0.0735} & 0.1444 & 0.0988 & 0.0914 \\
& SEAM & \textbf{0.0424} & 0.0851 & 0.0510 & 0.0497\\
\midrule
\multirow{3}{*}{\textbf{MSE $\downarrow$}} & Overthrust & \textbf{0.0057} & 0.0244 & 0.0166 & 0.0108 \\
& Marmousi 2 & \textbf{0.0161} & 0.0480 & 0.0290 & 0.0217 \\
& SEAM & \textbf{0.0043} & 0.0227 & 0.0082 & 0.0055 \\
\bottomrule[1.5pt]
\end{tabular}
\caption{The quantitative metrics of various methods on Overthrust, Marmousi 2 and SEAM datasets.}
\label{synthetic_metric}
\end{table}
\begin{table}[!htbp]
\centering
\tiny
\setlength{\tabcolsep}{2.5mm}
\renewcommand\arraystretch{1.6}
\begin{tabular}{cccccc}
\toprule[1.5pt]
\textbf{Metric} & \textbf{Dataset}  & \textbf{Proposed} & \textbf{\makecell{Semi-supervised \\ SM}} & \textbf{ADDIN} & \textbf{EIF} \\
\midrule
\multirow{3}{*}{\textbf{\makecell{SNR \\ (Mean $\uparrow$ $/$Std $\downarrow$)}}} & Overthrust & 33.2202$/$0.3101 & 27.6205$/$0.5682 & 29.2039$/$1.2417 & 30.3388$/$0.5423 \\
& Marmousi 2 & 26.6897$/$0.2826 & 22.5896$/$0.3135 & 23.9723$/$0.6895 & 25.2341$/$0.6262 \\
& SEAM & 33.0855$/$0.8053 & 25.1809$/$1.5377 & 27.1264$/$1.9302 & 30.8438$/$1.0290 \\
\midrule
\multirow{3}{*}{\textbf{\makecell{$R^2$ \\ (Mean $\uparrow$ $/$Std $\downarrow$)}}} & Overthrust & 0.9931$/$6.62$\times 10^{-4}$ & 0.9772$/$2.84$\times 10^{-3}$ & 0.9837$/$4.62$\times 10^{-3}$ & 0.9878$/$1.53$\times 10^{-3}$ \\
& Marmousi 2 & 0.9825$/$1.1390$\times 10^{-3}$ & 0.9550$/$3.35$\times 10^{-3}$ & 0.9669$/$5.30$\times 10^{-3}$ & 0.9753$/$3.66$\times 10^{-3}$ \\
& SEAM & 0.9960$/$7.91$\times 10^{-4}$ & 0.9726$/$6.26$\times 10^{-3}$ & 0.9831$/$7.83$\times 10^{-3}$ & 0.9933$/$1.58$\times 10^{-3}$ \\
\midrule
\multirow{3}{*}{\textbf{\makecell{SSIM \\ (Mean $\uparrow$ $/$Std $\downarrow$)}}} & Overthrust & 0.9678$/$2.32$\times 10^{-3}$ & 0.8987$/$7.91$\times 10^{-3}$ & 0.9355$/$1.52$\times 10^{-2}$ & 0.9531$/$1.74$\times 10^{-3}$\\
& Marmousi 2 & 0.9367$/$4.68$\times 10^{-3}$ & 0.8427$/$7.43$\times 10^{-3}$ & 0.8822$/$1.22$\times 10^{-2}$ & 0.9140$/$6.41$\times 10^{-3}$ \\
& SEAM & 0.9683$/$7.74$\times 10^{-3}$ & 0.8787$/$2.17$\times 10^{-2}$ & 0.9269$/$2.45$\times 10^{-2}$ & 9197$/$1.52$\times 10^{-2}$ \\
\midrule
\multirow{3}{*}{\textbf{\makecell{MAE \\ (Mean $\downarrow$ $/$Std $\downarrow$)}}} & Overthrust & 0.0478$/$2.30$\times 10^{-3}$ & 0.1063$/$5.12$\times 10^{-3}$ & 0.0670$/$1.35$\times 10^{-2}$ & 0.0603$/$3.28$\times 10^{-3}$\\
& Marmousi 2 & 0.0783$/$2.87$\times 10^{-3}$ & 0.1408$/$6.25$\times 10^{-3}$ & 0.1051$/$9.66$\times 10^{-3}$ & 0.0974$/$7.21$\times 10^{-3}$ \\
& SEAM & 0.0421$/$3.96$\times 10^{-3}$ & 0.0853$/$1.42$\times 10^{-2}$ & 0.0630$/$1.67$\times 10^{-2}$ & 0.0534$/$6.50$\times 10^{-3}$\\
\midrule
\multirow{3}{*}{\textbf{\makecell{MSE \\ (Mean $\downarrow$ $/$Std $\downarrow$)}}} & Overthrust & 0.0067$/$6.57$\times 10^{-4}$ & 0.0225$/$2.42$\times 10^{-3}$ & 0.0151$/$4.86$\times 10^{-3}$ & 0.0122$/$1.52$\times 10^{-3}$ \\
& Marmousi 2 & 0.0171$/$9.61$\times 10^{-4}$ & 0.0449$/$3.51$\times 10^{-3}$ & 0.03212$/$5.36$\times 10^{-3}$ & 0.0245$/$3.71$\times 10^{-3}$ \\
& SEAM & 0.0038$/$6.30$\times 10^{-4}$ & 0.0247$/$8.60$\times 10^{-3}$ & 0.0157$/$8.02$\times 10^{-3}$ & 0.0067$/$1.50$\times 10^{-3}$ \\
\bottomrule[1.5pt]
\end{tabular}
\caption{The mean and standard deviation of each metric under multiple sets of different random initializations.}
\label{synthetic_metric_mean_std}
\end{table}
Considering that these two metrics are influenced by the data range, we standardize $I$ and $I^{\prime}$ before calculating them. The metrics of each method across different datasets are presented in Table~\ref{synthetic_metric}. These metrics quantitatively characterize the performance disparities between the proposed method and the three comparative methods, and holistically reflect the advantages of the proposed method. To further measure the stability of each method, we initialized the models using multiple sets of random seeds and calculated the mean and standard deviation of each metric, as shown in Table~\ref{synthetic_metric_mean_std}. It can be observed that under different initializations (with different random seeds), the proposed method achieves the highest mean value for each metric, which demonstrates the effectiveness of the proposed learning strategy. Furthermore, it exhibits the smallest standard deviation for each metric, which indicates that the entire inversion process is the most stable and least affected by random initializations. As we mentioned before, the semi-supervised method such as ADDIN that adopts a data-driven forward model yields unstable results due to its involvement in the cascaded learning of two models. This instability is reflected in the larger standard deviations of each metric in Table~\ref{synthetic_metric_mean_std}, indicating that its results are more significantly affected by random initializations. 

\subsubsection{Ablation Experiment}

In addition to comparative experiments, ablation experiments can verify the effectiveness of the proposed improvements. We designed three control groups to compare and verify the effectiveness of the proposed physics-informed cross-learning framework, and the results of each group are presented in Table~\ref{ablation}. When only independent learning ($Loss_I$) is employed, the framework is equivalent to supervised training without the involvement of unlabeled data and WaveNet, which constitutes the simplest form of deep learning-based inversion framework. Consequently, inversion of unlabeled data becomes significantly challenging under this scenario. The integration of the WaveNet structure enables the calculation of $Loss_P$ and $Loss_C$. Physics-informed learning ($Loss_P$) introduces physical constraints, which effectively improves inversion accuracy. However, it still faces challenges in ensuring the inversion stability of unlabeled data, which is particularly pronounced on Overthrust and SEAM datasets. The combination of independent learning and cross-learning ($Loss_C$) yields results that are closest to those of the physics-informed cross-learning, with the primary performance gap observed specifically on the Overthrust dataset. Eventually, only the proposed method that integrates all three learning paradigms achieves the best and most stable results across all three datasets simultaneously. 

From the perspective of computational cost-effectiveness, the additional complexity introduced mainly comes from WaveNet, which accounts for approximately 50$\%$ of the computational complexity of the proposed method. The integration of $Loss_P$ and $Loss_C$ leads to a significant improvement in performance; meanwhile, the computational complexity of calculating $Loss_P$ and $Loss_C$ is relatively very low. Therefore, the integration of WaveNet is highly necessary.

\begin{table}[!htbp]
\centering
\tiny
\setlength{\tabcolsep}{3mm}
\renewcommand\arraystretch{1.6}
\begin{tabular}{cccccc}
\toprule[1.5pt]
\textbf{Metric} & \textbf{Dataset}  & \textbf{\makecell{$Loss_I+$ \\ $ Loss_C + Loss_P$}} & \textbf{$Loss_I$} & \textbf{$Loss_I+Loss_P$} & \textbf{$Loss_I+ Loss_C$} \\
\midrule
\multirow{3}{*}{\textbf{SNR $\uparrow $}} & Overthrust & \textbf{33.4297} & 25.1679 & 30.9230 & 32.0184 \\
& Marmousi 2 & \textbf{27.0205} & 21.0509 & 26.8535 & 26.5646 \\
& SEAM & \textbf{32.6203} & 28.7036 & 31.6426 & 32.4207 \\
\midrule
\multirow{3}{*}{\textbf{$R^2 \uparrow$}} & Overthrust & \textbf{0.9941} & 0.9602 & 0.9894 & 0.9918 \\
& Marmousi 2 & \textbf{0.9838} & 0.9360 & 0.9835 & 0.9827 \\
& SEAM & \textbf{0.9956} & 0.9892 & 0.9945 & 0.9955 \\
\midrule
\multirow{3}{*}{\textbf{SSIM $\uparrow $}} & Overthrust & \textbf{0.9728} & 0.8956 & 0.9549 & 0.9608\\
& Marmousi 2 & \textbf{0.9409} & 0.8323 & 0.9405 & 0.9401 \\
& SEAM & \textbf{0.9510} & 0.9270 & 0.9487 & 0.9506 \\
\midrule
\multirow{3}{*}{\textbf{MAE $\downarrow$}} & Overthrust & \textbf{0.0443} & 0.0977 & 0.0605 & 0.0491\\
& Marmousi 2 & \textbf{0.0735} & 0.1629 & 0.0767 & 0.0795 \\
& SEAM & \textbf{0.0424} & 0.0632 & 0.0443 & 0.0426 \\
\midrule
\multirow{3}{*}{\textbf{MSE $\downarrow$}} & Overthrust & \textbf{0.0057} & 0.0400 & 0.0106 & 0.0082 \\
& Marmousi 2 & \textbf{0.0161} & 0.0553 & 0.0163 & 0.0168 \\
& SEAM & \textbf{0.0043} & 0.0098 & 0.0049 & 0.0044 \\
\bottomrule[1.5pt]
\end{tabular}
\caption{The ablation experiments on Overthrust, Marmousi 2 and SEAM datasets.}
\label{ablation}
\end{table}

\subsection{Field Datasets}

\begin{figure}
\centering
\noindent\includegraphics[width=\textwidth]{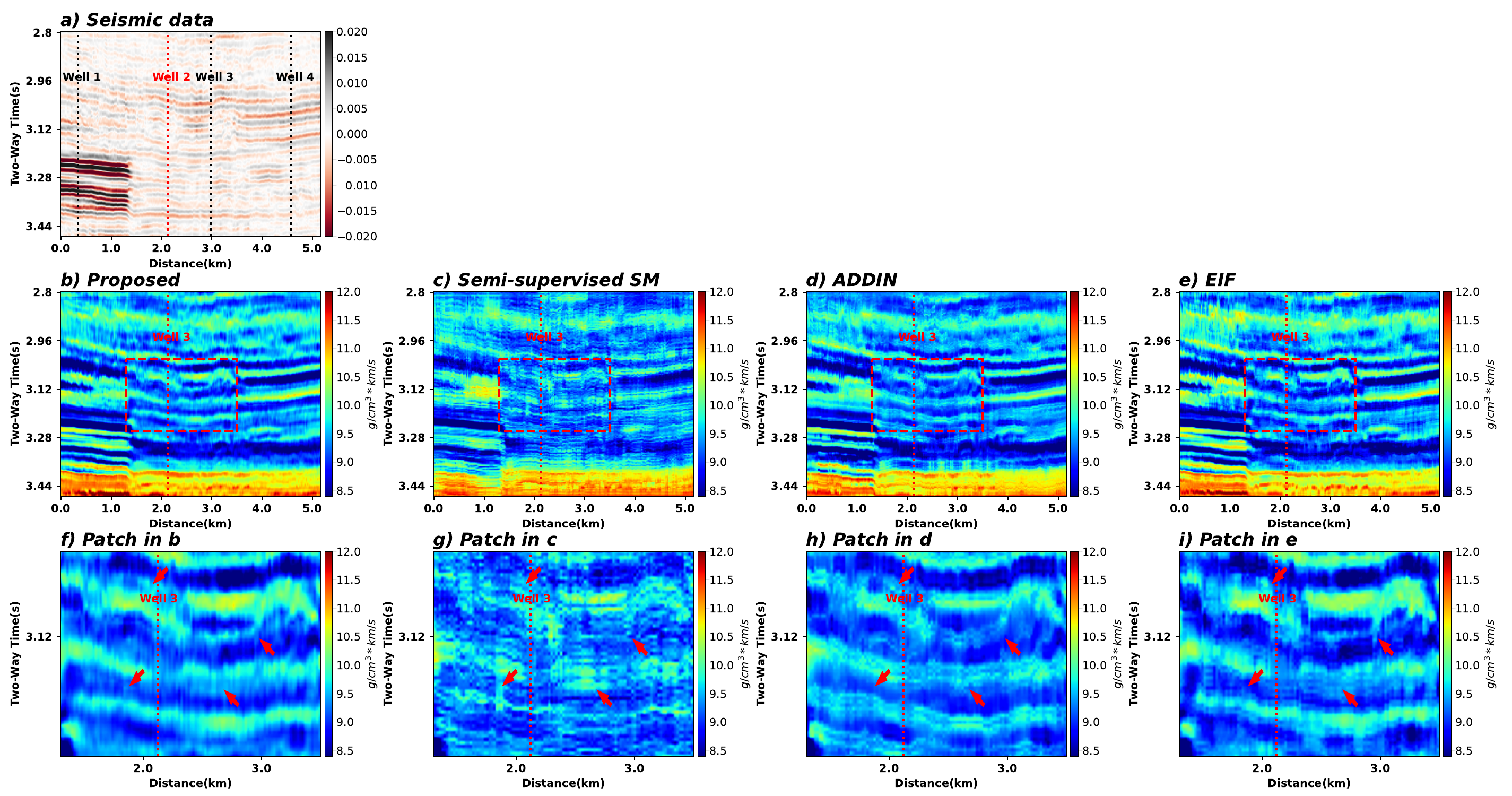}
\caption{The seismic data, wells and results of field data. a) represents field seismic data and wells; b) to e) represent the results of proposed method, semi-supervised SM, ADDIN, and EIF, respectively; f) to i) represent the zoom of the red boxed regions in b) to e), respectively.}
\label{field_whole}
\end{figure}

\begin{figure}
\centering
\noindent\includegraphics[width=\textwidth]{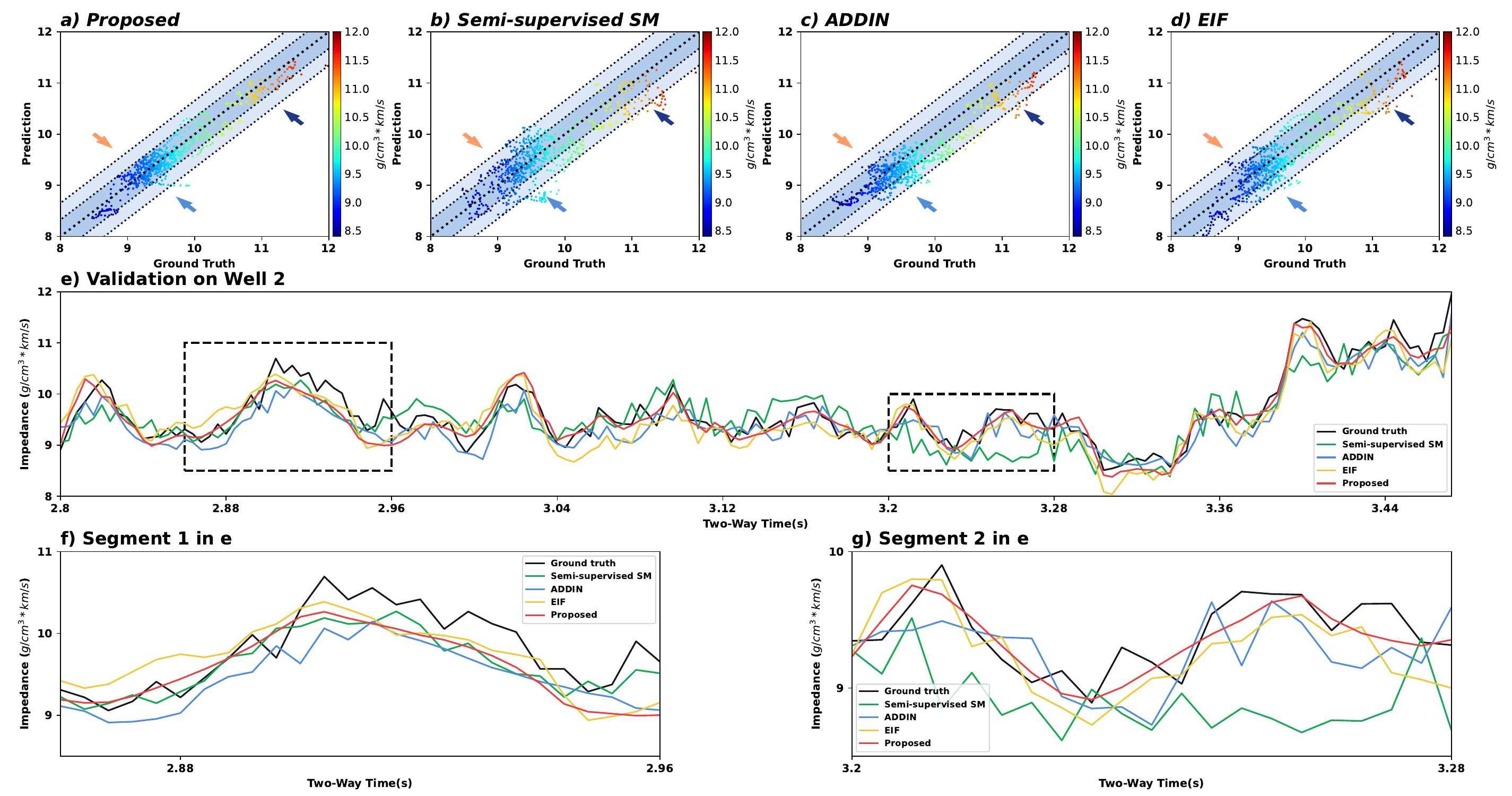}
\caption{The scatterplots and curves of results and Well 2. a) to d) represent the Well 2 scatterplots of proposed method, semi-supervised SM, ADDIN, and EIF, respectively; e) represents comparison of various methods on Well 2.}
\label{field_scatter_trace}
\end{figure}

Besides synthetic datasets, we further verified the practicality of the proposed method by utilizing field datasets. The field data used in this study were gathered from the Sichuan Basin in southwestern China, with a time interval of 4ms and a trace spacing of 25m. Sichuan basin is a continental sedimentary one, marked by multi-stage tectonic activities. As depicted in Figure~\ref{field_whole}a, the main target layers are composed of sandstone-mudstone interbeds. In these layers, tight gas reservoirs feature low porosity, low permeability, and high heterogeneity, along with frequent lateral and vertical lithofacies variations. There is a notable impedance contrast between sandstones and mudstones. Moreover, the region has undergone numerous tectonic events, resulting in the formation of faults and fractures. This not only renders seismic data more complex but also heightens the difficulty of inversion tasks. The field datasets contain four wells: three wells (Well 1, Well 3, and Well 4) were utilized for training, while one well (Well 2) was set aside for validation as a blind well (Figure~\ref{field_scatter_trace}). To ensure rigor, the parameter settings of all methods were kept consistent with those used in the synthetic dataset experiments.

From the perspective of overall inversion results, the impedance distribution of the proposed method exhibits clearer geological horizon continuity (Figure~\ref{field_whole}b). Particularly in the primary target layer segment with a two-way time of 2.96 to 3.28s, the boundary between high impedance (10 to 11 $g/cm^3*km/s$) and low impedance (8.5 to 9.5 $g/cm^3*km/s$) is more distinct. It can well match the variation trend of signals in the original seismic data, and the impedance transition smoothly in the lateral direction, with no obvious abrupt changes or impedance disorders at faults. This is consistent with the geological background of the Sichuan Basin, where significant lateral lithological variations are observed. By contrast, the overall inversion results of the three comparative methods exhibit obvious shortcomings: For the semi-supervised SM method, partial high-impedance regions become blurred in the 3.12 to 3.28s interval, making it difficult to distinguish the boundaries between adjacent sandstone reservoirs; In the ADDIN results, high-impedance zones are relatively scattered, leading to poor differentiation from low-impedance formations; As for the EIF method, it shows slight deficiencies in both lateral continuity and inversion accuracy for thin layers.

\begin{figure}
\centering
\noindent\includegraphics[width=\textwidth]{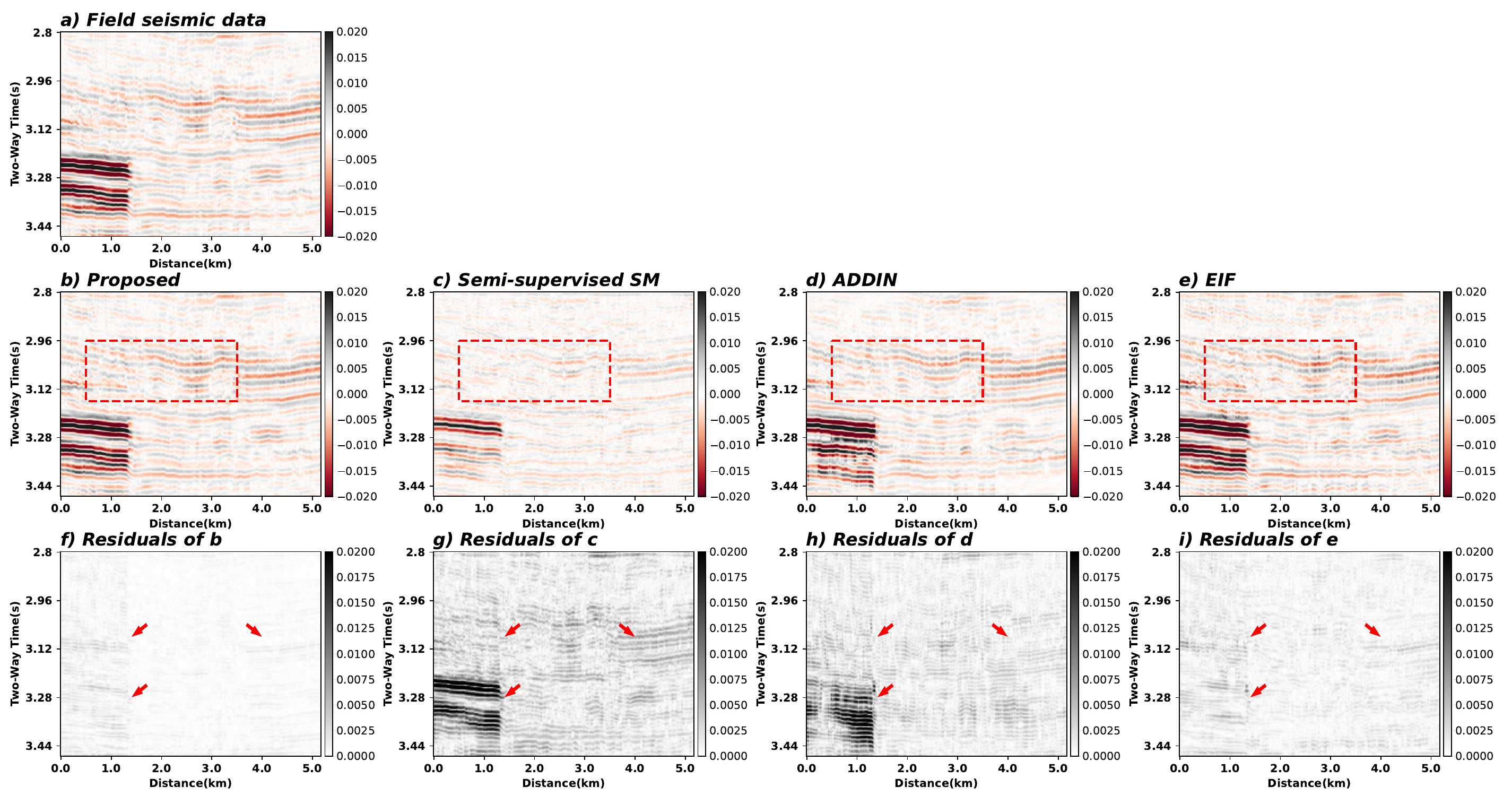}
\caption{The seismic data re-synthesized using the impedance results of each method, along with their absolute residuals relative to the field seismic data. }
\label{re-synthetic_field}
\end{figure}
\begin{table}[!htbp]
\centering
\tiny
\setlength{\tabcolsep}{1.2mm}
\renewcommand\arraystretch{1.6}
\begin{tabular}{ccc|cc|cc|cc}
\toprule[1.5pt]
\textbf{Metric} & \multicolumn{2}{c}{\textbf{Proposed}} & \multicolumn{2}{c}{\textbf{\makecell{Semi-supervised \\ SM}}} & \multicolumn{2}{c}{\textbf{ADDIN}} & \multicolumn{2}{c}{\textbf{EIF}}\\
\cline{2-9}
 & \makecell{re-synthesized \\ seismic data} & Well 2 & \makecell{re-synthesized \\ seismic data} & Well 2 & \makecell{re-synthesized \\ seismic data} & Well 2 & \makecell{re-synthesized \\ seismic data} & Well 2 \\
\midrule
\textbf{SNR $\uparrow $} & 18.1256 & 33.0727 & 2.2507 & 28.0027 & 4.2747 & 29.8023 & 12.7521 & 29.8846 \\
\textbf{$R^2 \uparrow$} & 0.9846 & 0.8962 & 0.4044 & 0.6664 & 0.6263 & 0.7796 & 0.9469 & 0.7837 \\
\textbf{SSIM $\uparrow $} & 0.9877 & $\backslash$ & 0.6998 & $\backslash$ & 0.8482 & $\backslash$ & 0.9450 & $\backslash$ \\
\textbf{MAE $\downarrow$} & 0.0892 & 0.0967 & 0.5696 & 0.3350 & 0.3675 & 0.1424 & 0.1745 & 0.1841 \\
\textbf{MSE $\downarrow$} & 0.0152 & 0.2397 & 0.7019 & 0.4598 & 0.4031 & 0.2957 & 0.0530 & 0.3476\\
\bottomrule[1.5pt]
\end{tabular}
\caption{The results of re-synthesized seismic data and blind well validation.}
\label{compare_re-synthetic_field}
\end{table}

Furthermore, we used the inversion results to re-synthesize seismic data and compared it with the field seismic data, as shown in Figure~\ref{re-synthetic_field} and Table~\ref{compare_re-synthetic_field}. It can be clearly observed that the semi-supervised SM and ADDIN methods exhibit significant resolution deficiencies in relatively complex interbedded zones, such as the region within 1km distance, which deficiency results in larger absolute errors of the synthesized seismic data in this region (Figures~\ref{re-synthetic_field}g and h). The proposed method and EIF perform relatively better, which is also clearly evident from the blind well validation (Table~\ref{compare_re-synthetic_field} and Figure~\ref{field_scatter_trace}). Although this value is lower than that obtained in the synthetic dataset experiments, considering various deviations under field conditions, the result is sufficiently satisfactory to support subsequent high-precision oil and gas development. Among all methods, semi-supervised SM yields the lowest metrics, which stems from its over-reliance on seismic wavelet extracted from wells and is susceptible to the quality of well data. ADDIN and EIF, though better, are hindered by the failure to integrate effective physical constraints with deep learning networks and the inherent complexity of field data. In future work, further evaluating and enhancing the field applicability of the proposed method is a direction worthy of exploration. 

\section{Discussion}
\subsection{Robustness Evaluation}

Beyond the comparative experiments mentioned above, verifying the robustness of the proposed method is equally crucial. In this section, we conduct tests under extreme scenarios, including different noise levels, various seismic wavelets, and varying numbers of wells, as shown in Table~\ref{noise-wavelet-well}. 
\begin{table}[!htbp]
\centering
\tiny
\setlength{\tabcolsep}{1.2mm}
\renewcommand\arraystretch{1.6}
\begin{tabular}{ccccc|ccc|ccc}
\toprule[1.5pt]
\textbf{Metric} & \textbf{Dataset} & \multicolumn{3}{c}{\textbf{SNR of Seismic Data}} & \multicolumn{3}{c}{\textbf{Wavelet Types}} & \multicolumn{3}{c}{\textbf{Well Nums}}\\
\cline{3-5}\cline{6-8}\cline{9-11}
 & & \textbf{5} & \textbf{10} & \textbf{15} & Ricker & Berlage & Generalized & 4 & 6 & 8 \\
\midrule
\multirow{3}{*}{\textbf{SNR $\uparrow $}} & Overthrust & 24.723 & 25.8145 & 30.2839 & 33.4297 & 37.1988 & 33.1309 & 28.2581 & 29.0782 & 29.8104\\
& Marmousi 2 & 18.3991 & 20.7711 & 23.8487 & 27.0205 & 29.6912 & 26.8863 & 21.4134 & 23.7617 & 24.5012\\
& SEAM & 23.9847 & 27.1374 & 29.1505 & 32.6203 & 36.7034 & 34.3375 & 22.8832 & 26.3399 & 28.1116\\
\midrule
\multirow{3}{*}{\textbf{$R^2 \uparrow$}} & Overthrust & 0.9559 & 0.9657 & 0.9877 & 0.9941 & 0.9975 & 0.9936 & 0.9804 & 0.9838 & 0.9843 \\
& Marmousi 2 & 0.8821 & 0.9317 & 0.9664 & 0.9838 & 0.9912 & 0.9833 & 0.9411 & 0.9657 & 0.9711\\
& SEAM & 0.9681 & 0.9845 & 0.9903 & 0.9956 & 0.9983 & 0.9971 & 0.9379 & 0.9883 & 0.9877\\
\midrule
\multirow{3}{*}{\textbf{SSIM $\uparrow $}} & Overthrust & 0.4868 & 0.6869 & 0.7889 & 0.9728 & 0.9880 & 0.9731 & 0.9556 & 0.9563 & 0.9567\\
& Marmousi 2 & 0.3843 & 0.5603 & 0.7124 & 0.9409 & 0.9694 & 0.9397 & 0.9115 & 0.9071 & 0.9203\\
& SEAM & 0.4724 & 0.6036 & 0.7387 & 0.9510 & 0.9729 & 0.9629 & 0.9120 & 0.9161 & 0.9585\\
\midrule
\multirow{3}{*}{\textbf{MAE $\downarrow$}} & Overthrust & 0.1612 & 0.1268 & 0.0800 & 0.0443 & 0.0331 & 0.0449 & 0.0834 & 0.0775 & 0.0700\\
& Marmousi 2 & 0.2565 & 0.1898 & 0.1216 & 0.0735 & 0.0580 & 0.0803 & 0.1743 & 0.1289 & 0.1038\\
& SEAM & 0.1340 & 0.0926 & 0.0722 & 0.0424 & 0.0266 & 0.0339 & 0.1571 & 0.0686 & 0.0673\\
\midrule
\multirow{3}{*}{\textbf{MSE $\downarrow$}} & Overthrust & 0.0445 & 0.0346 & 0.0119 & 0.0057 & 0.0024 & 0.0064 & 0.0193 & 0.0156 & 0.0135\\
& Marmousi 2 & 0.1177 & 0.0684 & 0.0307 & 0.0161 & 0.0087 & 0.0166 & 0.0594 & 0.0346 & 0.0289\\
& SEAM &0.0320 & 0.0154 & 0.0096 & 0.0043 & 0.0016 & 0.0029 & 0.0563 & 0.0116 & 0.0070\\
\bottomrule[1.5pt]
\end{tabular}
\caption{The results of the proposed method on each dataset under different seismic data noise levels, different wavelet types, and different numbers of wells.}
\label{noise-wavelet-well}
\end{table}

From the perspective of noise levels, we added random noise to the seismic data to generate three datasets with different SNR as inputs for inversion. It can be observed that noise exerts a significant impact on the proposed method, and all evaluation metrics show a noticeable decline, particularly under high noise levels (SNR$=$5). In field exploration, with the advancement of acquisition technologies, such cases are generally rare. First, the AVO stacking technique enhances the SNR of the data; second, this issue can be mitigated by a variety of noise attenuation techniques \citep{23}. It is undeniable that the proposed method is susceptible to noise, which merits further research in future work.

On the other hand, in addition to the most commonly used Ricker wavelet, we conducted experiments with Berlage \citep{48} and Generalized \citep{49} wavelets. The type of seismic wavelet also exerts a notable impact on inversion results: the performance with the Berlage wavelet is significantly superior to that with the other two wavelets. This is attributed to the simpler structure of the Berlage wavelet compared to the Ricker and Generalized wavelets. However, in practical applications, seismic wavelets are often more complex. In future work, we plan to address this issue by incorporating wavelet constraints into the training process, aiming to further enhance inverison performance. Finally, we conducted more extreme experiments with a reduced number of wells. It can be observed that as the number of wells decreases, all evaluation metrics show a downward trend. Notably, in exploration work involving large-scale and complex geological structures (such as those of the Marmousi 2 and SEAM datasets), it is unreasonable to rely on an extremely small number of wells. Nevertheless, the proposed method still achieves a certain level of effectiveness. Effectively performing acoustic inversion with a small number of wells remains one of the most challenging problems in the field.

\begin{table}[!htbp]
\centering
\tiny
\setlength{\tabcolsep}{1.2mm}
\renewcommand\arraystretch{1.6}
\begin{tabular}{ccccccc}
\toprule[1.5pt]
 \multicolumn{2}{c}{\textbf{Metric}} & \textbf{Proposed Methods}  & \textbf{\makecell{Semi-supervised \\ SM}} & \textbf{ADDIN} & \textbf{EIF}\\
\midrule
\multirow{4}{*}{\textbf{Time(s) $\downarrow$}} & \textbf{Overthrust} & 55 & 58 & 71 & 30 \\
& \textbf{Marmousi 2} & 86 & 112 & 141 & 51 \\
& \textbf{SEAM} & 83 & 125 & 157 & 68 \\
& \textbf{Field data} & 21 & 17 & 49 & 9 \\
\bottomrule[1.5pt]
\end{tabular}
\caption{The time consumption of each method.}
\label{efficiency}
\end{table}

In addition, computational efficiency is also a crucial metric. Under unified hardware (a Intel Core i5-14400F CPU with 16GB memory and a Nvidia RTX 4060 with 8GB memory) and software conditions, we statistically analyzed the time consumed by each method in processing each dataset, as presented in Table~\ref{efficiency}. It can be observed that the proposed method maintains relatively balanced computational efficiency while ensuring inversion accuracy, although it is lower than that of the EIF method. Further improving computational efficiency will facilitate the inversion work of large-scale exploration projects and represents a direction worthy of research in the future.

\subsection{Wavelet Evaluation}

\begin{figure}
\centering
\noindent\includegraphics[width=\textwidth]{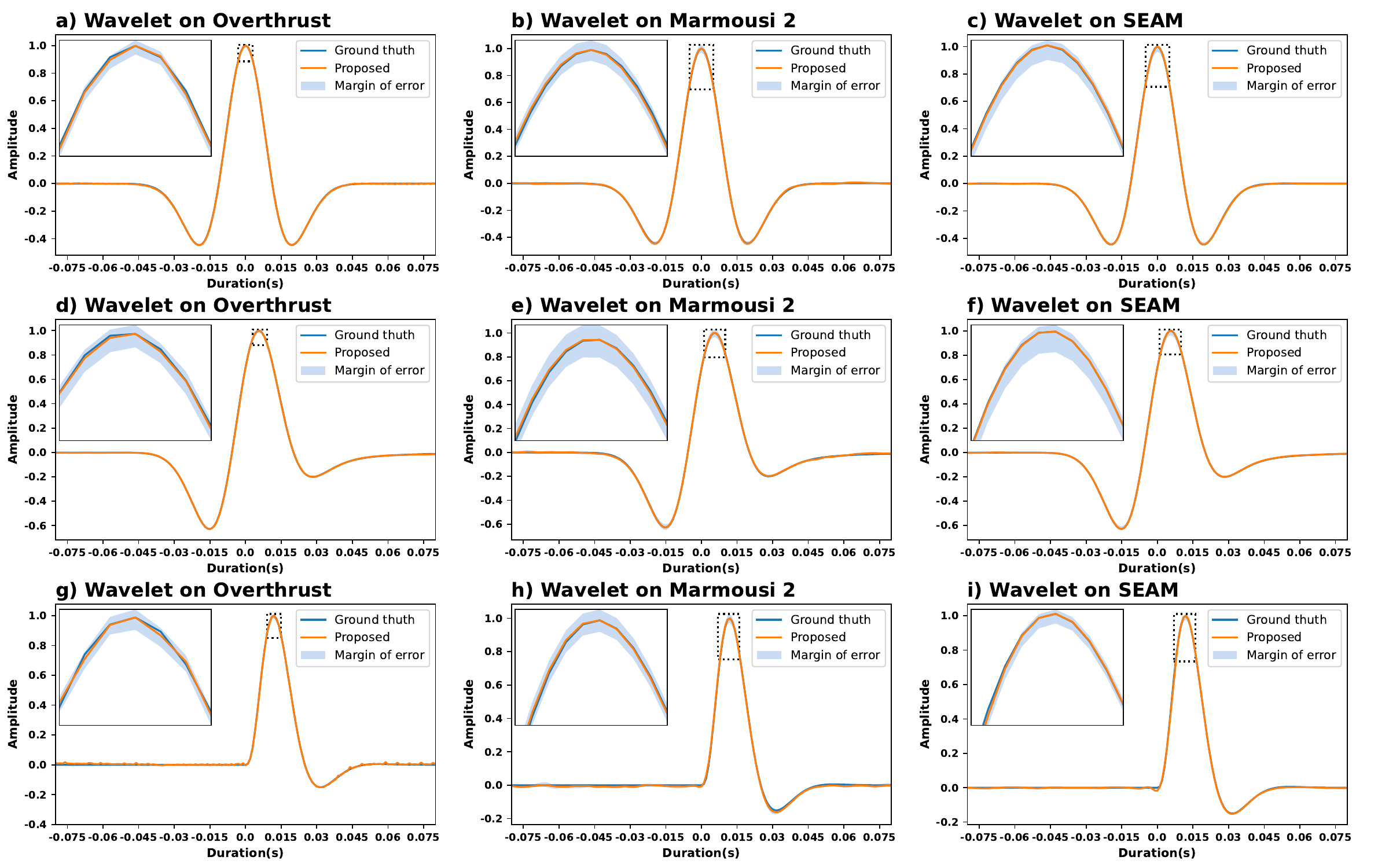}
\caption{The wavelets extracted from the synthetic datasets. a) to c) represent Ricker wavelet, d) to f) represent generalized wavelet, and g) to i) represent Berlage wavelet.}
\label{synthetic_wavelet}
\end{figure}

\begin{table}[!htbp]
\centering
\tiny
\setlength{\tabcolsep}{1.2mm}
\renewcommand\arraystretch{1.6}
\begin{tabular}{cccccc|cccc|cccc}
\toprule[1.5pt]
 & \textbf{Dataset} & \multicolumn{4}{c}{\textbf{Overthrust}} & \multicolumn{4}{c}{\textbf{Marmousi 2}} & \multicolumn{4}{c}{\textbf{SEAM}}\\
\cline{3-14}
 & \textbf{Freq(Hz)} & \textbf{15} & \textbf{20} & \textbf{25} & \textbf{30} & \textbf{15} & \textbf{20} & \textbf{25} & \textbf{30} & \textbf{15} & \textbf{20} & \textbf{25} & \textbf{30} \\
\midrule
\multirow{5}{*}{\textbf{Phase}} & \textbf{$+0$} & 0.9987 & 0.9988 & 0.9987 & 0.9997 & 0.9988 & 0.9987 & 0.9988 & 0.9992 & 0.9987 & 0.9988 & 0.9984 & 0.9997 \\
& \textbf{$+0.2\pi$} & 0.9985 & 0.9987 & 0.9988 & 0.9996 & 0.9984 & 0.9986 & 0.9987 & 0.9991 & 0.9988 & 0.9985 & 0.9987 & 0.9996\\
& \textbf{$+0.4\pi$} & 0.9982 & 0.9985 & 0.9988 & 0.9998 & 0.9985 & 0.9985 & 0.9985 & 0.9992 & 0.9988 & 0.9987 & 0.9988 & 0.9997\\
& \textbf{$+0.6\pi$} & 0.9983 & 0.9983 & 0.9985 & 0.9997 & 0.9986 & 0.9984 & 0.9983 & 0.9991 & 0.9987 & 0.9989 & 0.9987 & 0.9995\\
& \textbf{$+0.8\pi$} & 0.9982 & 0.9981 & 0.9988 & 0.9998 & 0.9987 & 0.9986 & 0.9984 & 0.9993 & 0.9988 & 0.9988 & 0.9988 & 0.9996\\
\bottomrule[1.5pt]
\end{tabular}
\caption{The $R^2$ between extracted wavelet and ground truth with different dominant frequency and phases.}
\label{wavelet_r2}
\end{table}

\begin{figure}
\centering
\noindent\includegraphics[width=\textwidth]{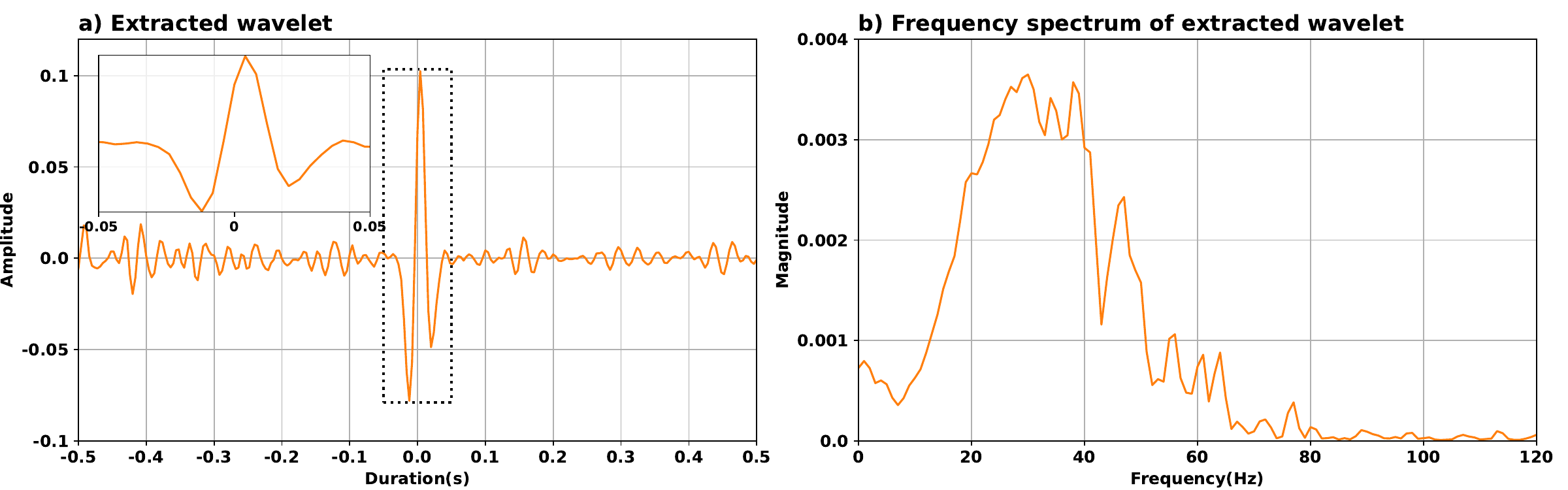}
\caption{The wavelets extracted and its frequency spectrum from the field datasets.}
\label{field_wavelet}
\end{figure}
Although the primary objective in inversion work is acoustic impedance, the wavelets synchronously extracted in the proposed method are also worthy of evaluation to verify the rationality of the proposed method. First, we plotted the wavelets extracted from the synthetic datasets, as shown in Figure~\ref{synthetic_wavelet}. It can be observed that the proposed method is capable of extracting different wavelets effectively, with a small discrepancy between these extracted wavelets and the real wavelets used for synthesizing seismic data, which provides positive constraints for acoustic impedance inversion. 

In addition to the zero-phase wavelet (Ricker wavelet), mixed-phase wavelet (Generalized wavelet) and minimum-phase wavelet (Berlage wavelet), the dominant frequency and phase of the wavelets are usually uncertain. Therefore, we conducted further experiments to evaluate the WaveNet. Since Ricker wavelets are the most commonly used zero-phase wavelets, we simulated wavelets with different phases by rotating the phase of Ricker wavelets, and the $R^2$ of results are presented in Table~\ref{wavelet_r2}. It can be observed that WaveNet exhibits low sensitivity to both the phase and dominant frequency of wavelets while maintaining high overall accuracy. We hypothesize that this is because wavelets have a simpler structure compared to impedance and their frequency bandwidth is limited; consequently, the prediction accuracy of WaveNet for wavelets is higher than that of ImpedanceNet for impedance. The accuracy of WaveNet's prediction results is beneficial for both the training of ImpedanceNet and the entire inversion process.

In field scenarios, limitations imposed by discrepancies in data acquisition and processing can complicate both the inversion and wavelet extraction tasks. In Figure~\ref{field_wavelet}, we have plotted the seismic wavelets extracted from field data. It can be observed that the wavelets exhibit significant complexity over the entire time range. From the frequency-magnitude spectrum, it can be observed that the wavelet energy is concentrated in the range of 20 to 40Hz overall. Additionally, the wavelets exhibit a relatively low dominant frequency and a narrow bandwidth, which increases the difficulty of inversion work. Nevertheless, the extracted wavelets generally conform to the expectations for field data, exhibiting distinct strong amplitudes within a short duration. The main reasons for the significant complexity of wavelets in field data may stem from the following aspects: First, post-stack data is typically used in acoustic impedance inversion, which is obtained through extensive preprocessing of pre-stack data. The preprocessing of pre-stack data as well as well-logging data (such as amplitude variation with offset stack, time-depth matching, and seismic signal processing) is usually complex and involves subjective factors and easily introduce certain errors. Second, well data in field scenarios are localized, meaning there exist inherent discrepancies between different well data. These challenges are commonly encountered in all acoustic impedance inversion work, and the integration of prior knowledge or the implementation of wavelet constraints (for instance, wavelets typically exist within an extremely short duration) represents a direction worthy of further exploration in future research.

\subsection{Weights of Losses}

In this work, the proposed physics-informed cross-learning involves a loss function consisting of three components: $Loss_I$, $Loss_P$, and $Loss_C$, where $Loss_I$ and $Loss_C$ contain parts that optimize ImpedanceNet and WaveNet, respectively. Therefore, as described in Equations~\ref{loss_I} and~\ref{loss_C}, we adopted equal weights for each component in $Loss_I$ and $Loss_C$. However, a question remains regarding how to select the weights among $Loss_I$, $Loss_P$, and $Loss_C$ during the training process. By referring to various multi-task learning strategies \citep{52}, we conducted comparative experiments using uncertainty weighting (UW) \citep{53} in this section, which can be specifically expressed as:
\begin{equation}
\begin{split}
    Loss(\theta_E, \theta_W, \theta_I, \sigma_1, \sigma_2,\sigma_3) = \frac{1}{2\sigma_1^2}Loss_I + \frac{1}{2\sigma_2^2}Loss_P \\ + \frac{1}{2\sigma_3^2}Loss_C + \sum_{i=1}^{3}\log\sigma_i
\end{split}
\end{equation}
\begin{table}[!htbp]
\centering
\tiny
\setlength{\tabcolsep}{1.2mm}
\renewcommand\arraystretch{1.6}
\begin{tabular}{cccc|cc|cc}
\toprule[1.5pt]
\textbf{Metric} & \textbf{Dataset} & \multicolumn{2}{c}{\textbf{Ricker}} & \multicolumn{2}{c}{\textbf{Berlage}} & \multicolumn{2}{c}{\textbf{Generalized}}\\
\cline{3-5}\cline{6-8}
 & & Equal weights & UW & Equal weights & UW & Equal weights & UW \\
\midrule
\multirow{3}{*}{\textbf{SNR $\uparrow $}} & Overthrust & 33.4297 & \textbf{33.6644}
 & \textbf{37.1988} & 32.2935 & \textbf{33.1309} & 30.9824\\
& Marmousi 2 & \textbf{27.0205} & 26.5281 & \textbf{29.6912} & 28.8538 & \textbf{26.8863} & 26.6147\\
& SEAM & 32.6203 & \textbf{33.6988} & \textbf{36.7034} & 36.2228 & 34.3375 & \textbf{35.2010} \\
\midrule
\multirow{3}{*}{\textbf{$R^2 \uparrow$}} & Overthrust & 0.9941 & \textbf{0.9944} & \textbf{0.9975} & 0.9923 & \textbf{0.9936} & 0.9896 \\
& Marmousi 2 & \textbf{0.9838} & 0.9819 & \textbf{0.9912} & 0.9894 & \textbf{0.9833} & 0.9822\\
& SEAM & 0.9956 & \textbf{0.9966} & \textbf{0.9983} & 0.9981 & 0.9971 & \textbf{0.9977}\\
\midrule
\multirow{3}{*}{\textbf{SSIM $\uparrow $}} & Overthrust & 0.9728 & \textbf{0.9742} & \textbf{0.9880} & 0.9687 & \textbf{0.9731} & 0.9678\\
& Marmousi 2 & \textbf{0.9409} & 0.9258 & \textbf{0.9694} & 0.9647 & \textbf{0.9397} & 0.9276\\
& SEAM & 0.9510 & \textbf{0.9587} & 0.9729 & \textbf{0.9847} & 0.9629 & \textbf{0.9635}\\
\midrule
\multirow{3}{*}{\textbf{MAE $\downarrow$}} & Overthrust & 0.0443 & \textbf{0.0410} & \textbf{0.0331} & 0.0558 & \textbf{0.0449} & 0.0648\\
& Marmousi 2 & \textbf{0.0735} & 0.0828 & \textbf{0.0580} & 0.0628 & 0.0803 & \textbf{0.0798}\\
& SEAM & 0.0424 & \textbf{0.0360} & \textbf{0.0266} & 0.0285 & 0.0339 & \textbf{0.0315}\\
\midrule
\multirow{3}{*}{\textbf{MSE $\downarrow$}} & Overthrust & 0.0057 & \textbf{0.0054} & \textbf{0.0024} & 0.0074 & \textbf{0.0064} & 0.0103\\
& Marmousi 2 & \textbf{0.0161} & 0.0171 & \textbf{0.0087} & 0.0096 & \textbf{0.0166} & 0.0170\\
& SEAM & 0.0043 & \textbf{0.0029} & \textbf{0.0016} & 0.0019 & 0.0029 & \textbf{0.0021}\\
\bottomrule[1.5pt]
\end{tabular}
\caption{Comparison of equal weights and UW.}
\label{compare_UW}
\end{table}
\begin{figure}
\centering
\noindent\includegraphics[width=\textwidth]{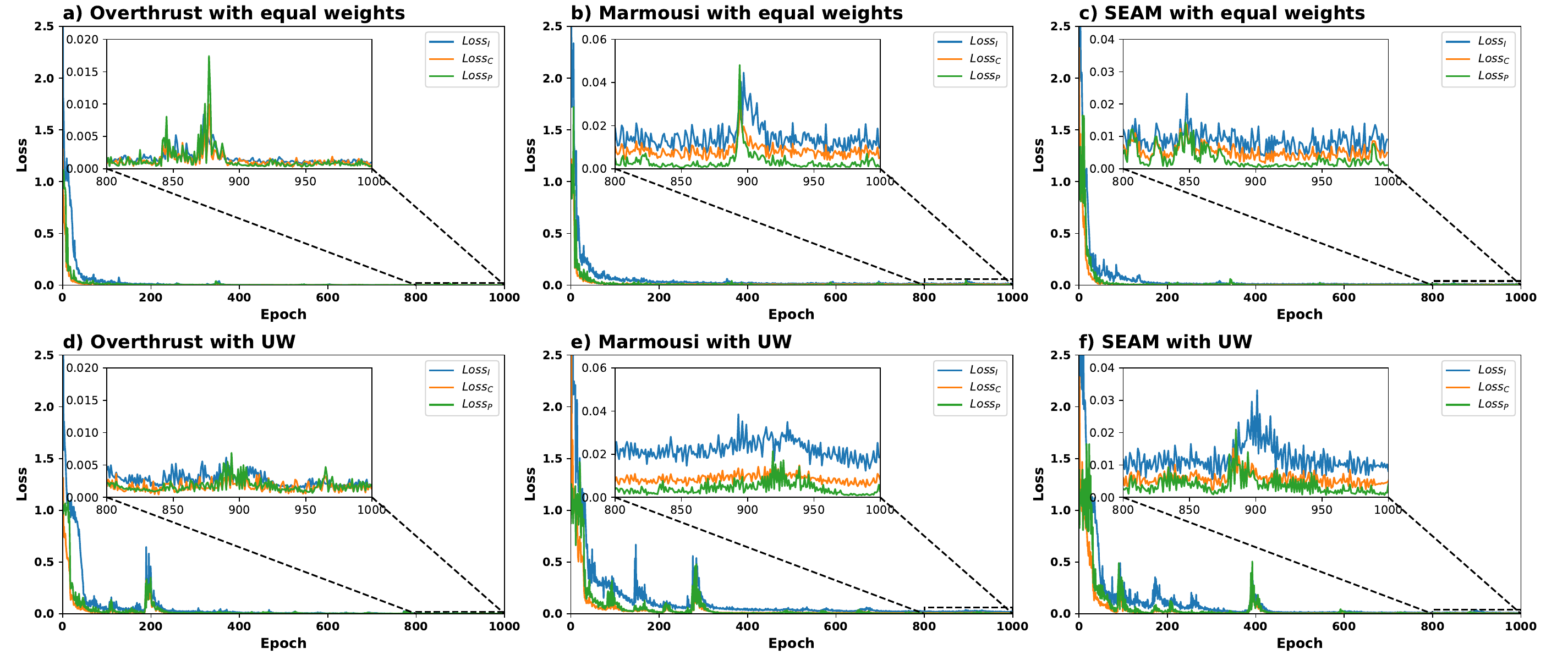}
\caption{The loss cruves of equal weights and UW with ricker wavelet.}
\label{compare_UW_ricker}
\end{figure}
and the results of the UW are shown in Table~\ref{compare_UW}. Overall, the gap between equal weights and UW is small, and UW only achieves better results in some cases. We believe the reason for this phenomenon is that there is strong complementarity among the three proposed loss functions, whereas in other multi-task learning works, the correlation between different tasks is relatively weak. On the other hand, loss weights can be used to balance losses of different scales \citep{54}. In this work, we applied Z-Score standardization to the data and all three loss functions adopted mean squared error to ensure that the scales of the three loss functions are consistent. Finally, we analyzed the convergence during the training process for different weight strategies, with the Ricker wavelet taken as an example (Figure~\ref{compare_UW_ricker}). It can be observed that both equal weights and UW eventually achieve good convergence. In comparison, equal weights exhibit local fluctuations (Figures~\ref{compare_UW_ricker}a and b), which we attribute to adjustments to local solutions; UW does not show such a phenomenon, but its final convergence value is higher than that of equal weights (Figures~\ref{compare_UW_ricker}e and f).

In summary, adjusting the weights of different losses is beneficial in some cases and the UW version will be provided in our open-source code. However, simply calculating a weighted sum at the loss level is insufficient, as it can lead to unstable training and results. A promising solution is to use multi-objective optimization \citep{55, 56}; however, this will significantly reduce the overall computational efficiency while only bringing limited improvement in performance. 

\section{Conclusion}

In this work, we designed a brand-new deep learning framework tailored to the characteristics of seismic acoustic impedance inversion, and integrated the physical forward process into both the framework structure and training process. At the framework structure level, we designed a framework that includes an encoder and a downstream collaborative network, which is used for simultaneous impedance inversion and seismic wavelet extraction. At the training strategy level, considering that seismic acoustic impedance inversion is a few-shot problem, the training strategy we proposed consists of three components, simultaneously incorporating physical constraints (physics-informed learning), label-supervised learning (independent learning), and domain-adaptive learning (cross-learning). We refer to the proposed method as a physics-informed cross-learning framework and tested it on three widely used synthetic datasets as well as one complex field dataset. The results demonstrate that the proposed method is more stable and effective compared with previous semi-supervised methods, and the extracted wavelets achieve high accuracy. Finally, the codes of proposed method will be made publicly available to support future research.  

\section{Acknowledgments}

This research is supported by the Chengdu University of Technology Postgraduate Innovative Cultivation Program. The code and examples of this work can be found in \url{https://github.com/lexiaoheng/Physics-Informed-Cross-Learning-for-Seismic-Acoustic-Impedance-Inversion-and-Wavelet-Extraction} and long-term maintenance version will be uploaded to a GitHub repository \url{https://github.com/lexiaoheng/Mariana}.

\bibliography{cite}

@book{1,
  title={Seismic reservoir characterization: An earth modelling perspective},
  author={Doyen, Philippe M},
  year={2007},
  publisher={EAGE}
}

@article{2,
  title={Time-frequency mixed domain multi-trace simultaneous inversion method},
  author={Lin, Kai and Zhao, Lian and Wen, Xiaotao and Zhang, Yuqiang},
  journal={Geoenergy Science and Engineering},
  volume={230},
  pages={212164},
  year={2023},
  publisher={Elsevier}
}

@article{3,
  title={Data-driven inverse modelling through neural network (deep learning) and computational heat transfer},
  author={Tamaddon-Jahromi, Hamid Reza and Chakshu, Neeraj Kavan and Sazonov, Igor and Evans, Llion M and Thomas, Hywel and Nithiarasu, Perumal},
  journal={Computer Methods in Applied Mechanics and Engineering},
  volume={369},
  pages={113217},
  year={2020},
  publisher={Elsevier}
}

@article{4,
  title={A review of least-squares inversion and its application to geophysical problems},
  author={Lines, LR and Treitel, S},
  journal={Geophysical prospecting},
  volume={32},
  number={2},
  pages={159--186},
  year={1984},
  publisher={Wiley Online Library}
}

@article{5,
  title={Fractal-based stochastic inversion of poststack seismic data using very fast simulated annealing},
  author={Srivastava, RP and Sen, MK},
  journal={Journal of Geophysics and Engineering},
  volume={6},
  number={4},
  pages={412--425},
  year={2009},
  publisher={Oxford University Press}
}

@article{6,
  title={Fast seismic inversion methods using ant colony optimization algorithm},
  author={Conti, Cassio Rodrigo and Roisenberg, Mauro and Neto, Guenther Schwedersky and Porsani, Milton Jos{\'e}},
  journal={IEEE Geoscience and Remote Sensing Letters},
  volume={10},
  number={5},
  pages={1119--1123},
  year={2013},
  publisher={IEEE}
}

@article{7,
  title={Model-based inversion of amplitude-variations-with-offset data using a genetic algorithm},
  author={Mallick, Subhashis},
  journal={Geophysics},
  volume={60},
  number={4},
  pages={939--954},
  year={1995},
  publisher={Society of Exploration Geophysicists}
}

@article{8,
  title={Particle swarm optimization: A new tool to invert geophysical data},
  author={Shaw, Ranjit and Srivastava, Shalivahan},
  journal={Geophysics},
  volume={72},
  number={2},
  pages={F75--F83},
  year={2007},
  publisher={Society of Exploration Geophysicists}
}

@article{9,
  title={Bayesian seismic inversion based on rock-physics prior modeling for the joint estimation of acoustic impedance, porosity and lithofacies},
  author={de Figueiredo, Leandro Passos and Grana, Dario and Santos, Marcio and Figueiredo, Wagner and Roisenberg, Mauro and Neto, Guenther Schwedersky},
  journal={Journal of Computational Physics},
  volume={336},
  pages={128--142},
  year={2017},
  publisher={Elsevier}
}

@article{10,
  title={Alternate iterative inversion of acoustic impedance and wavelet based on sparse constraints},
  author={Zhao, Lian and Cao, Danping and Zhang, Yuqiang and An, Zhidi and Lin, Kai and Wen, Xiaotao},
  journal={IEEE Transactions on Geoscience and Remote Sensing},
  year={2025},
  publisher={IEEE}
}

@book{11,
  title={Seismic inversion: Theory and applications},
  author={Wang, Yanghua},
  year={2016},
  publisher={John Wiley \& Sons}
}

@article{12,
  title={Stable and efficient seismic impedance inversion using quantum annealing with L1 norm regularization},
  author={Wang, Silin and Liu, Cai and Li, Peng and Chen, Changle and Song, Chao},
  journal={Journal of Geophysics and Engineering},
  volume={21},
  number={1},
  pages={330--343},
  year={2024},
  publisher={Oxford University Press}
}

@article{13,
  title={A comprehensive survey of few-shot learning: Evolution, applications, challenges, and opportunities},
  author={Song, Yisheng and Wang, Ting and Cai, Puyu and Mondal, Subrota K and Sahoo, Jyoti Prakash},
  journal={ACM Computing Surveys},
  volume={55},
  number={13s},
  pages={1--40},
  year={2023},
  publisher={ACM New York, NY}
}

@article{14,
  title={Convolutional neural network for seismic impedance inversion},
  author={Das, Vishal and Pollack, Ahinoam and Wollner, Uri and Mukerji, Tapan},
  journal={Geophysics},
  volume={84},
  number={6},
  pages={R869--R880},
  year={2019},
  publisher={Society of Exploration Geophysicists}
}

@incollection{15,
  title={Estimation of acoustic impedance from seismic data using temporal convolutional network},
  author={Mustafa, Ahmad and Alfarraj, Motaz and AlRegib, Ghassan},
  booktitle={SEG technical program expanded abstracts 2019},
  pages={2554--2558},
  year={2019},
  publisher={Society of Exploration Geophysicists}
}

@article{16,
  title={Seismic impedance inversion based on residual attention network},
  author={Wu, Bangyu and Xie, Qiao and Wu, Baohai},
  journal={IEEE Transactions on Geoscience and Remote Sensing},
  volume={60},
  pages={1--17},
  year={2022},
  publisher={IEEE}
}

@article{17,
  title={TransInver: 3D data-driven seismic inversion based on self-attention},
  author={Li, Kewen and Dou, Yimin and Xiao, Yuan and Jing, Ruilin and Zhu, Jianbing and Ma, Chengjie},
  journal={Geophysics},
  volume={89},
  number={1},
  pages={WA127--WA141},
  year={2024},
  publisher={Society of Exploration Geophysicists}
}

@article{18,
  title={Semisupervised sequence modeling for elastic impedance inversion},
  author={Alfarraj, Motaz and AlRegib, Ghassan},
  journal={Interpretation},
  volume={7},
  number={3},
  pages={SE237--SE249},
  year={2019},
  publisher={Society of Exploration Geophysicists and American Association of Petroleum~…}
}

@article{19,
  title={Double-scale supervised inversion with a data-driven forward model for low-frequency impedance recovery},
  author={Yuan, Sanyi and Jiao, Xinqi and Luo, Yaneng and Sang, Wenjing and Wang, Shangxu},
  journal={Geophysics},
  volume={87},
  number={2},
  pages={R165--R181},
  year={2022},
  publisher={Society of Exploration Geophysicists}
}

@article{20,
  title={Semi-supervised learning seismic inversion based on spatio-temporal sequence residual modeling neural network},
  author={Song, Lei and Yin, Xingyao and Zong, Zhaoyun and Jiang, Man},
  journal={Journal of Petroleum Science and Engineering},
  volume={208},
  pages={109549},
  year={2022},
  publisher={Elsevier}
}

@article{21,
  title={Nash-multitask learning-semisupervised temporal convolutional network method for prestack three-parameter inversion},
  author={Liu, Yingtian and Li, Yong and Li, Huating and Peng, Junheng and Liao, Zhangquan and Feng, Wen and Wang, Mingwei},
  journal={Geophysics},
  volume={90},
  number={4},
  pages={R175--R193},
  year={2025},
  publisher={Society of Exploration Geophysicists}
}

@article{23,
  title={Fast diffusion model for seismic data noise attenuation},
  author={Peng, Junheng and Li, Yong and Liu, Yingtian and Wang, Mingwei and Liao, Zhangquan and Wang, Xiaowen},
  journal={Geophysics},
  volume={90},
  number={4},
  pages={1--55},
  year={2025},
  publisher={Society of Exploration Geophysicists}
}

@article{24,
  title={FaultVitNet: A Vision Transformer Assisted Network for 3D Fault Segmentation},
  author={Li, Chao and Fomel, Sergey and Chen, Yangkang and Dommisse, Robin and Savvaidis, Alexandros},
  journal={Journal of Geophysical Research: Machine Learning and Computation},
  volume={2},
  number={2},
  pages={e2024JH000488},
  year={2025},
  publisher={Wiley Online Library}
}

@article{25,
  title={Seismic Data Reconstruction via Least Squares Generative Adversarial Networks with Inverse Interpolation},
  author={Li, Chao and Liu, Xingye and Zu, Shaohuan},
  journal={IEEE Transactions on Geoscience and Remote Sensing},
  year={2025},
  publisher={IEEE}
}

@article{27,
  title={Seismic impedance inversion based on deep learning with geophysical constraints},
  author={Su, Yuqi and Cao, Danping and Liu, Shiyou and Hou, Zhiyu and Feng, Jihao},
  journal={Geoenergy Science and Engineering},
  volume={225},
  pages={211671},
  year={2023},
  publisher={Elsevier}
}

@article{29,
  title={Porosity prediction using semi-supervised learning with biased well log data for improving estimation accuracy and reducing prediction uncertainty},
  author={Sang, Wenjing and Yuan, Sanyi and Han, Hongwei and Liu, Haojie and Yu, Yue},
  journal={Geophysical Journal International},
  volume={232},
  number={2},
  pages={940--957},
  year={2023},
  publisher={Oxford University Press}
}

@article{30,
  title={Acoustic impedance prediction using an attention-based dual-branch double-inversion network},
  author={Feng, Wen and Liu, Yingtian and Li, Yong and Li, Huating and Wang, Xiaowen},
  journal={Earth Science Informatics},
  volume={18},
  number={1},
  pages={1--20},
  year={2025},
  publisher={Springer}
}

@article{31,
  title={Iterative Gradient Corrected Semi-Supervised Seismic Impedance Inversion via Swin Transformer},
  author={Pang, Qi and Chen, Hongling and Gao, Jinghuai and Wang, Zhiqiang and Yang, Ping},
  journal={IEEE Transactions on Geoscience and Remote Sensing},
  year={2025},
  publisher={IEEE}
}

@article{32,
  title={Seismic inversion based on fusion neural network for the joint estimation of acoustic impedance and porosity},
  author={Sun, Hui and Zhang, Jian and Xue, Yiran and Zhao, Xiaoyan},
  journal={IEEE Transactions on Geoscience and Remote Sensing},
  year={2024},
  publisher={IEEE}
}

@article{33,
  title={High-resolution closed-loop seismic inversion network in the time-frequency-phase mixed domain},
  author={Liu, Yingtian and Li, Yong and Peng, Junheng and Wang, Mingwei and Wang, Xiaowen and Xie, Jianyong},
  journal={Geophysics},
  volume={90},
  number={6},
  pages={M271--M288},
  year={2025},
  publisher={Society of Exploration Geophysicists}
}

@article{34,
  title={Impedance inversion by using the low-frequency full-waveform inversion result as an a priori model},
  author={Yuan, Sanyi and Wang, Shangxu and Luo, Yaneng and Wei, Wanwan and Wang, Guanchao},
  journal={Geophysics},
  volume={84},
  number={2},
  pages={R149--R164},
  year={2019},
  publisher={Society of Exploration Geophysicists}
}

@article{35,
  title={Predictive decomposition of time series with application to seismic exploration},
  author={Robinson, Enders A},
  journal={Geophysics},
  volume={32},
  number={3},
  pages={418--484},
  year={1967},
  publisher={Society of Exploration Geophysicists}
}

@article{36,
  title={Seismic data strong noise attenuation based on diffusion model and principal component analysis},
  author={Peng, Junheng and Li, Yong and Liao, Zhangquan and Wang, Xuben and Yang, Xingyu},
  journal={IEEE Transactions on Geoscience and Remote Sensing},
  volume={62},
  pages={1--11},
  year={2024},
  publisher={IEEE}
}

@inproceedings{37,
  title={A comparative study of transfer learning methodologies and causality for seismic inversion with temporal convolutional networks},
  author={Mustafa, Ahmad and AlRegib, Ghassan},
  booktitle={SEG International Exposition and Annual Meeting},
  pages={D011S067R001},
  year={2021},
  organization={SEG}
}

@article{38,
  title={Robust deep learning-based seismic inversion workflow using temporal convolutional networks},
  author={Smith, Robert and Nivlet, Philippe and Alfayez, Hussain and AlBinHassan, Nasher},
  journal={Interpretation},
  volume={10},
  number={2},
  pages={SC41--SC55},
  year={2022},
  publisher={Society of Exploration Geophysicists and American Association of Petroleum~…}
}

@article{39,
  title={Low-Frequency Model Prediction of Acoustic Impedance Based on Spatio-Temporal Gated Recurrent Unit Fusion Network},
  author={Dai, Ruiqi and Cao, Danping and Liu, Qiang},
  journal={IEEE Transactions on Geoscience and Remote Sensing},
  year={2025},
  publisher={IEEE}
}

@article{40,
  title={Seismic Porosity Prediction via Semi-Supervised Learning: Integrating a Low-Frequency Model and a Closed-Loop Network Structure},
  author={Chen, Yuanyuan and Zhao, Luanxiao and Liu, Jingyu and Huang, He and Geng, Jianhua},
  journal={IEEE Transactions on Geoscience and Remote Sensing},
  year={2025},
  publisher={IEEE}
}

@inproceedings{41,
  title={Open set domain adaptation},
  author={Panareda Busto, Pau and Gall, Juergen},
  booktitle={Proceedings of the IEEE international conference on computer vision},
  pages={754--763},
  year={2017}
}

@article{42,
  title={Domain-adversarial training of neural networks},
  author={Ganin, Yaroslav and Ustinova, Evgeniya and Ajakan, Hana and Germain, Pascal and Larochelle, Hugo and Laviolette, Fran{\c{c}}ois and March, Mario and Lempitsky, Victor},
  journal={Journal of machine learning research},
  volume={17},
  number={59},
  pages={1--35},
  year={2016}
}

@article{43,
  title={Open set domain adaptation: Theoretical bound and algorithm},
  author={Fang, Zhen and Lu, Jie and Liu, Feng and Xuan, Junyu and Zhang, Guangquan},
  journal={IEEE transactions on neural networks and learning systems},
  volume={32},
  number={10},
  pages={4309--4322},
  year={2020},
  publisher={IEEE}
}

@article{44,
  title={Marmousi2: An elastic upgrade for Marmousi},
  author={Martin, Gary S and Wiley, Robert and Marfurt, Kurt J},
  journal={The leading edge},
  volume={25},
  number={2},
  pages={156--166},
  year={2006},
  publisher={Society of Exploration Geophysicists}
}

@article{45,
  title={Three dimensional SEG/EAEG models—An update},
  author={Aminzadeh, Fred and Burkhard, N and Long, J and Kunz, Tim and Duclos, P},
  journal={The Leading Edge},
  volume={15},
  number={2},
  pages={131--134},
  year={1996},
  publisher={Society of Exploration Geophysicists}
}

@article{46,
  title={SEAM: A 10-year trail of success grounded in sound research, collaboration, and outreach},
  author={Capello, Maria A and House, Nancy and Barkhouse, William},
  journal={The Leading Edge},
  volume={36},
  number={9},
  pages={724--727},
  year={2017},
  publisher={Society of Exploration Geophysicists Tulsa, Oklahoma}
}

@article{47,
  title={Encoder-Inverter Framework for Seismic Acoustic Impedance Inversion},
  author={Peng, Junheng and Liu, Yingtian and Wang, Mingwei and Li, Yong and Feng, Wen},
  journal={arXiv preprint arXiv:2507.19933},
  year={2025}
}

@article{48,
  title={The berlage wavelet},
  author={Aldridge, David F},
  journal={Geophysics},
  volume={55},
  number={11},
  pages={1508--1511},
  year={1990},
  publisher={Society of Exploration Geophysicists}
}

@article{49,
  title={Generalized seismic wavelets},
  author={Wang, Yanghua},
  journal={Geophysical Journal International},
  volume={203},
  number={2},
  pages={1172--1178},
  year={2015},
  publisher={Oxford University Press}
}

@article{50,
  title={3D seismic mask auto encoder: Seismic inversion using transformer-based reconstruction representation learning},
  author={Dou, Yimin and Li, Kewen},
  journal={Computers and Geotechnics},
  volume={169},
  pages={106194},
  year={2024},
  publisher={Elsevier}
}

@article{51,
  title={Toward more-robust, AI-enabled subsurface seismic imaging for geotechnical applications},
  author={Vantassel, Joseph P and Bhochhibhoya, Sanish},
  journal={Computers and Geotechnics},
  volume={187},
  pages={107443},
  year={2025},
  publisher={Elsevier}
}

@article{52,
  title={Multi-task learning for dense prediction tasks: A survey},
  author={Vandenhende, Simon and Georgoulis, Stamatios and Van Gansbeke, Wouter and Proesmans, Marc and Dai, Dengxin and Van Gool, Luc},
  journal={IEEE transactions on pattern analysis and machine intelligence},
  volume={44},
  number={7},
  pages={3614--3633},
  year={2021},
  publisher={IEEE}
}

@inproceedings{53,
  title={Multi-task learning using uncertainty to weigh losses for scene geometry and semantics},
  author={Kendall, Alex and Gal, Yarin and Cipolla, Roberto},
  booktitle={Proceedings of the IEEE conference on computer vision and pattern recognition},
  pages={7482--7491},
  year={2018}
}

@inproceedings{54,
  title={Gradnorm: Gradient normalization for adaptive loss balancing in deep multitask networks},
  author={Chen, Zhao and Badrinarayanan, Vijay and Lee, Chen-Yu and Rabinovich, Andrew},
  booktitle={International conference on machine learning},
  pages={794--803},
  year={2018},
  organization={PMLR}
}

@article{55,
  title={Multi-task learning as multi-objective optimization},
  author={Sener, Ozan and Koltun, Vladlen},
  journal={Advances in neural information processing systems},
  volume={31},
  year={2018}
}

@article{56,
  title={Pareto multi-task learning},
  author={Lin, Xi and Zhen, Hui-Ling and Li, Zhenhua and Zhang, Qing-Fu and Kwong, Sam},
  journal={Advances in neural information processing systems},
  volume={32},
  year={2019}
}

@article{57,
  title={Data-efficient operator learning via unsupervised pretraining and in-context learning},
  author={Chen, Wuyang and Song, Jialin and Ren, Pu and Subramanian, Shashank and Morozov, Dmitriy and Mahoney, Michael W},
  journal={Advances in Neural Information Processing Systems},
  volume={37},
  pages={6213--6245},
  year={2024}
}
\bibliographystyle{elsarticle-harv}

\end{document}